  \pdfoutput=1
  \documentclass[graphicx,amssymb,amsmath,enumerate,11pt]{report}
  \usepackage{fancyhdr}
  \fancyheadoffset[]{0.1cm}

  \usepackage{epsfig}
  \usepackage{axodraw}
  \usepackage{url}
  \newcommand\mchapter[2]{\chapter*{#1}
  \vskip -0.5cm \noindent {\it \LARGE #2}
  \addcontentsline{toc}{chapter}{#1\\{\normalsize\it #2}}}

\usepackage{slashed}
\usepackage{multirow} 

\newcommand{\be}{\begin{equation}}
\newcommand{\ee}{\end{equation}}
\newcommand{\bea}{\begin{eqnarray}}
\newcommand{\eea}{\end{eqnarray}}

\newcommand{\eqn}[1]{eq.~(\ref{#1})}
\newcommand{\eqns}[2]{eqs.~(\ref{#1}) and (\ref{#2})}

\def \lsim{\mathrel{\vcenter
     {\hbox{$<$}\nointerlineskip\hbox{$\sim$}}}}
\def \gsim{\mathrel{\vcenter
     {\hbox{$>$}\nointerlineskip\hbox{$\sim$}}}}

\def \lesssim{\mathrel{\vcenter
     {\hbox{$<$}\nointerlineskip\hbox{$\sim$}}}}
\def \gtrsim{\mathrel{\vcenter
     {\hbox{$>$}\nointerlineskip\hbox{$\sim$}}}}

\def\gappeq{\mathrel{\rlap {\raise.5ex\hbox{$>$}}
{\lower.5ex\hbox{$\sim$}}}}
\def\lappeq{\mathrel{\rlap{\raise.5ex\hbox{$<$}}
{\lower.5ex\hbox{$\sim$}}}}

\newcommand{\Eqn}[1]{Eq.~(\ref{#1})}

\begin{document}      

 \rhead{\bfseries Leptogenesis in the Universe}

 \mchapter{Leptogenesis in the Universe}
 {Authors:\ Chee Sheng Fong$^a$, Enrico Nardi$^a$, Antonio Riotto$^b$}
 \label{ch-08:leptogenesis}

\vspace{0.5cm}

\begin{center}
$^a$ {\it INFN - Laboratori Nazionali di Frascati, 
                Via Enrico Fermi 40, 00044 Frascati, Italy} \\ [6pt]
$^b$ {\it Department of Theoretical Physics and Center for 
Astroparticle Physics (CAP), \\
University of Geneva, 24 quai E. Ansermet, CH-1211 Geneva 4, Switzerland}
\end{center}

\vspace{3cm}

\begin{center}
{\bf Abstract}
\end{center}
Leptogenesis is a class of scenarios in which the cosmic baryon
asymmetry originates from an initial lepton asymmetry generated in the
decays of heavy sterile neutrinos in the early Universe. We explain
why leptogenesis is an appealing mechanism for baryogenesis.  We
review its motivations, the basic ingredients, and describe subclasses
of effects, like those of lepton flavours, spectator processes,
scatterings, finite temperature corrections, the role of the heavier
sterile neutrinos and quantum corrections.
We then address leptogenesis in supersymmetric scenarios, as well as some 
other popular variations of the basic leptogenesis framework.


\itemsep -5pt
 \tableofcontents


\section{The Baryon Asymmetry of the Universe}
\label{sec-08:intro}

\subsection{Observations}
\label{sec-08:observations}
Up to date no traces of cosmological antimatter have been observed.
The presence of a small amount of antiprotons and positrons in cosmic
rays can be consistently explained by their secondary origin in cosmic
particles collisions or in highly energetic astrophysical processes,
but no antinuclei, even as light as anti-deuterium or as tightly
bounded as anti-$\alpha$ particles, has ever been detected.

The absence of annihilation radiation $p\bar p\to \dots \pi^0\to \dots
2\gamma$ excludes significant matter-antimatter admixtures in objects
up to the size of galactic clusters $\sim
20\,$Mpc~\cite{Steigman:1976ev}. Observational limits on
anomalous contributions to the cosmic diffuse $\gamma$-ray background
and the absence of distortions in the cosmic microwave background
(CMB) implies that little antimatter is to be found within
$\sim 1\,$Gpc and that within our horizon an equal amount of matter
and antimatter can be excluded~\cite{Cohen:1997ac}. Of course,
at larger super-horizon scales the vanishing of the average asymmetry
cannot be ruled out, and this would indeed be the case if the
fundamental Lagrangian is $C$ and $CP$ symmetric and charge invariance
is broken spontaneously~\cite{Dolgov:1991fr}.


Quantitatively, the value of baryon asymmetry of the Universe is
inferred from observations in two independent ways.  The first way is
by confronting the abundances of the light elements, $D$, $^3{\rm
  He}$, $^4{\rm He}$, and $^7${\rm Li}, with the predictions of Big
Bang nucleosynthesis
(BBN)~\cite{Iocco:2008va,Steigman:2007xt,Nakamura:2010zzi,%
  Steigman:2005uz,Cyburt:2004yc,Olive:1999ij}. The crucial time for
primordial nucleosynthesis is when the thermal bath temperature falls
below $T\lsim 1\,$MeV.  With the assumption of only three light
neutrinos, these predictions depend on a single parameter, that is the
difference between the number of baryons and anti-baryons normalized
to the number of photons:
\be
\eta\equiv \frac{n_B- n_{\bar{B}}}{n_{\gamma }} {\Big |}_{0}, 
\label{eq-08:etaB}
\ee
where the subscript $0$ means ``at present time''.  By using only
the abundance of deuterium, that is particularly sensitive to $\eta$,
Ref.~\cite{Iocco:2008va} quotes:
\begin{equation}
  \label{eq-08:2H}
10^{10}\,\eta = 5.7\pm 0.6\quad \qquad\qquad(95\%\> {\rm c.l.}) \,.
\end{equation}
In this same range there is also an acceptable agreement among the
various abundances, once theoretical uncertainties as well as
statistical and systematic errors are accounted for~\cite{Nakamura:2010zzi}.

The second way is from measurements of the CMB anisotropies (for
pedagogical reviews, see Refs.~\cite{Hu:2001bc,Dodelson:2003ft}).  The
crucial time for CMB is that of recombination, when the temperature
dropped down to $T\lsim 1\,$eV and neutral hydrogen can be formed.
CMB observations measure the relative baryon contribution to the
energy density of the Universe multiplied by the square of the
(reduced) Hubble constant $h\equiv H_0/(100\,{\rm km}\>{\rm
  sec}^{-1}\,{\rm Mpc}^{-1})$:
\begin{equation}
\label{eq-08:Omegab}
\Omega_Bh^2\equiv h^2\frac{\rho_B}{\rho_{\rm crit}}\,,
\end{equation}
that is related to $\eta$ through $10^{10}\eta=274\; \Omega_B\ h^2$.
The physical effect of the baryons at the onset of matter domination,
which occurs quite close to the recombination epoch, is to provide
extra gravity which enhances the compression into potential wells. The
consequence is enhancement of the compressional phases which
translates into enhancement of the odd peaks in the spectrum. Thus, a
measurement of the odd/even peak disparity constrains the baryon
energy density. A fit to the most recent observations (WMAP7 data
only, assuming a $\Lambda$CDM model with a scale-free power spectrum
for the primordial density fluctuations) gives at 68\%
c.l.~\cite{Larson:2010gs}
\begin{equation}
\label{eq-08:omecmb}
10^2\, \Omega_B h^2 = 2.258^{+0.057}_{-0.056}\,.
\end{equation}
There is a third way to express the baryon asymmetry of the Universe,
that is by normalizing the baryon asymmetry to the entropy density $s
=g_*(2\pi^2/45)T^3$, where $g_*$ is the number of degrees of freedom
in the plasma, and $T$ is the temperature:
\be
Y_{\Delta B} \equiv  \frac{n_B- n_{\bar{B}}}{s} {\Big |}_{0}\,. 
\label{eq-08:YB}
\ee
The relation with the previous definitions is given by the
conversion factor $s_0/n_{\gamma0}= 7.04$.  $Y_{\Delta B}$ is a
convenient quantity in theoretical studies of the generation of the
baryon asymmetry from very early times, because it is conserved
throughout the thermal evolution of the Universe.
In terms of $Y_{\Delta B}$ the BBN results \eqn{eq-08:2H} and the CMB
measurement \eqn{eq-08:omecmb} (at $95\% c.l.$)  read:
\begin{equation}
\label{eq-08:YB_CMB}
Y_{\Delta B}^{BBN} =  (8.10 \pm 0.85) \times 10^{-11},
\qquad 
Y_{\Delta B}^{CMB}=(8.79 \pm 0.44) \times 10^{-11}. 
\end{equation}
The impressive consistency between the determinations of the baryon
density of the Universe from BBN and CMB that, besides being
completely independent, also refer to epochs with a six orders of
magnitude difference in temperature, provides a striking confirmation
of the hot Big Bang cosmology.

\subsection{Theory}
\label{sec-08:theory}

From the theoretical point of view, the question is where the Universe
baryon asymmetry comes from.  
The inflationary cosmological model excludes the possibility of a fine
tuned initial condition, and since we do not know any other way to
construct a consistent cosmology without inflation, this is a strong veto.

The alternative possibility is that the Universe baryon asymmetry is
generated dynamically, a scenario that is known as {\em
  baryogenesis}. This requires that baryon number ($B$) is not conserved.
More precisely, as Sakharov pointed out~\cite{Sakharov:1967dj}, the
ingredients required for baryogenesis are three:

\begin{enumerate} \itemsep 1pt
\item $B$ violation is required to evolve from an initial state with
  $Y_{\Delta B}=0$ to a state with $Y_{\Delta B}\neq0$.
\item C and CP violation:
  If either C or CP were conserved, then processes involving baryons
  would proceed at the same rate as the C- or CP-conjugate
  processes involving antibaryons, with the overall effect that no
  baryon asymmetry is generated.
\item Out of equilibrium dynamics:
Equilibrium distribution functions $n_{\rm eq}$  are determined 
solely by the particle energy $E$, chemical potential $\mu$, and by 
 its mass which, because of the CPT
theorem, is the same for particles and antiparticles. 
%
%
When charges (such as $B$) are not conserved, the corresponding
chemical potentials vanish, and thus 
$n_B = \int \frac{d^3p}{(2\pi^3)} n_{\rm eq}=n_{\bar B}$. 
%
%
\end{enumerate}
Although these ingredients are all present in the Standard Model (SM),
so far all attempts to reproduce quantitatively the observed baryon
asymmetry have failed.
\begin{enumerate}  \itemsep 1pt
\item In the SM $B$ is violated by the triangle anomaly.  Although at
  zero temperature $B$ violating processes are too suppressed to have
  any observable effect~\cite{'tHooft:1976up}, at high temperatures
  they occur with unsuppressed rates~\cite{Kuzmin:1985mm}.  The
  first condition is then quantitatively realized in the early Universe.

\item SM weak interactions violate C maximally. However, the amount of
  CP violation from the Kobayashi-Maskawa complex
  phase~\cite{Kobayashi:1973fv}, as quantified by means of the Jarlskog
  invariant\cite{Jarlskog:1985ht}, is only of order $10^{-20}$, and this 
  renders impossible generating $Y_{\Delta
    B}\sim10^{-10}$~\cite{Gavela:1994ds,Gavela:1994dt,Huet:1994jb}.

\item Departures from thermal equilibrium occur in the SM at the
  electroweak phase transition (EWPT)~\cite{Rubakov:1996vz,Trodden:1998ym}.
However, the experimental lower bound on the Higgs mass implies 
that this transition is not sufficiently first order as
  required for successful baryogenesis~\cite{Kajantie:1995kf}.
\end{enumerate}
This shows that baryogenesis requires new physics that extends the SM
in at least two ways: It must introduce new sources of CP violation
and it must either provide a departure from thermal equilibrium in
addition to the EWPT or modify the EWPT itself.  In the past thirty
years or so, several new physics mechanisms for baryogenesis have been
put forth. Some among the most studied are {\it GUT
  baryogenesis}~\cite{Ignatiev:1978uf,Yoshimura:1978ex,%
  Toussaint:1978br,Dimopoulos:1978kv,Ellis:1978xg,Weinberg:1979bt,%
  Yoshimura:1979gy,Barr:1979ye,Nanopoulos:1979gx,Yildiz:1979gx}, {\it
  Electroweak baryogenesis}~\cite{Rubakov:1996vz,Riotto:1999yt,%
  Cline:2006ts}, the {\it Affleck-Dine
  mechanism}~\cite{Affleck:1984fy,Dine:1995kz}, {\it Spontaneous
  Baryogenesis}~\cite{Cohen:1987vi,Cohen:1988kt}. However, soon after
the discovery of neutrino masses, because of its connections with the
seesaw model~\cite{Minkowski:1977sc,%
  Yanagida:1979as,Glashow,GellMann:1980vs,Mohapatra:1980yp} and its
deep interrelations with neutrino physics in general, the mechanism of
baryogenesis via {\it Leptogenesis} acquired a continuously increasing
popularity. Leptogenesis was first proposed by Fukugita and Yanagida
in Ref.~\cite{Fukugita:1986hr}. Its simplest and theoretically best
motivated realization is precisely within the seesaw mechanism.  To
implement the seesaw, new Majorana $SU(2)_L$ singlet neutrinos with a
large mass scale $M$ are added to the SM particle spectrum.  The
complex Yukawa couplings of these new particles provide new sources of
CP violation, departure from thermal equilibrium can occur if their
lifetime is not much shorter than the age of the Universe when $T\sim
M$, and their Majorana masses imply that lepton number is not
conserved. A lepton asymmetry can then be generated dynamically, and
SM sphalerons will partially convert it into a baryon
asymmetry~\cite{Khlebnikov:1988sr}. A particularly interesting
possibility is ``thermal leptogenesis'' where the heavy Majorana
neutrinos are produced by scatterings in the thermal bath starting
from a vanishing initial abundance, so that their number density can
be calculated solely in terms of the seesaw parameters and of the
reheat temperature of the Universe.


This review is organized as follows: in Section~\ref{sec-08:basic} the
basis of leptogenesis are reviewed in the simple scenario of the one
flavour regime, while the role of flavour effects is described in
Section~\ref{sec-08:flavour}. Theoretical improvements of the basic
pictures, like spectator effects, scatterings and CP violation in
scatterings, thermal corrections, the possible role of the heavier
singlet neutrinos, and quantum effects are reviewed in
Section~\ref{sec-08:improving}.  Leptogenesis in the supersymmetric
seesaw is reviewed in Section~\ref{sec-08:susy}, while in
Section~\ref{sec-08:beyond} we mention possible leptogenesis
realizations that go beyond the type-I seesaw. Finally, in
Section~\ref{sec-08:conclusions} we draw the conclusions.



\vspace{1cm}
\section{$N_1$ Leptogenesis in the Single Flavour Regime}
\label{sec-08:basic}

The aim of this section is to give a pedagogical introduction to
leptogenesis~\cite{Fukugita:1986hr} and establish the notations. We will
consider the classic example of leptogenesis from the lightest
right-handed (RH) neutrino $N_1$ (the so-called $N_1$ leptogenesis) in
the type-I seesaw
model~\cite{Minkowski:1977sc,GellMann:1980vs,Yanagida:1979as,Mohapatra:1980yp}
in the single flavour regime. First in Section \ref{sec-08:typeI} we
introduce the type-I seesaw Lagrangian and the relevant parameters. In
Section \ref{sec-08:CPasy}, we will review the CP violation in RH
neutrino decays induced at 1-loop level. Then in Section
\ref{sec-08:CBE}, we will write down the classical Boltzmann equations
taking into account of only decays and inverse decays of $N_1$ and
give a simple but rather accurate analytical estimate of the solution.
In Section \ref{sec-08:LtoB} we will relate the lepton asymmetry
generated to the baryon asymmetry of the Universe.  Finally in Section
\ref{sec-08:DI}, we will discuss the lower bound on $N_1$ mass and the
upper bound on light neutrino mass scale from successful leptogenesis.

\subsection{Type-I seesaw, neutrino masses and leptogenesis}
\label{sec-08:typeI}

With $m\,(m \geq 2)$\footnote{Neutrino oscillation data and leptogenesis both require $m \geq 2$.} 
singlet RH neutrinos $N_{R_i}\,(i=1,m)$, 
we can add the following Standard Model (SM) gauge invariant terms to the SM Lagrangian
\bea
{\cal L}_{I} = {\cal L}_{SM} + i\overline{N_{R_i}} \slashed\partial N_{R_i} 
- \left( \frac{1}{2} M_i \overline {N^c_{R_i}} N_{R_i} 
+ \epsilon_{ab}Y_{\alpha i} \overline {N_{R_i}} \ell_\alpha^a H^{b} 
+ h.c.\right),
\label{eq-08:lag_I}
\eea
where $M_i$ are the Majorana masses of the RH neutrinos, 
$\ell_\alpha=(\nu_{\alpha L}, \alpha_L^-)$ with $\alpha=e,\mu,\tau$ and $H=(H^+,H^0)$ 
are respectively the left-handed (LH) lepton and Higgs $SU(2)_L$ doublets and
$\epsilon_{ab}=-\epsilon_{ba}$ with $\epsilon_{12}=1$. 
Without loss of generality, we have chosen the basis where the Majorana mass term is diagonal.
The physical mass eigenstates of the RH neutrinos are the Majorana neutrinos 
$N_i = N_{R_i} + N_{R_i}^c$. Since $N_i$ are gauge singlets, 
the scale of $M_i$ is naturally much larger 
than the electroweak (EW) scale $M_i \gg \langle\Phi\rangle \equiv v = 174$ GeV. 
Hence after EW symmetry breaking, the light neutrino mass matrix is given by 
the famous seesaw relation~\cite{Minkowski:1977sc,GellMann:1980vs,Yanagida:1979as,Mohapatra:1980yp}
\be
m_{\nu} \simeq -v^2 Y \frac{1}{M} Y^T.
\label{eq-08:seesaw}
\ee
Assuming $Y \sim {\cal O}(1)$ and $m_{\nu} \simeq \sqrt{\Delta m_{atm}^2} \simeq 0.05$ eV,
we have $M \sim 10^{15}$ GeV not far below the GUT scale.

Besides giving a natural explanation of the light neutrino masses,
there is another bonus: 
the \emph{three} Sakharov's conditions\cite{Sakharov:1967dj} for 
leptogenesis are implicit in \eqn{eq-08:lag_I}
with the \emph{lepton number violation} provided by $M_i$, 
the \emph{CP-violation} from the complexity
of $Y_{i\alpha}$ and the \emph{departure from thermal equilibrium condition} given
by an additional requirement that $N_i$ decay rate $\Gamma_{N_i}$ is not very fast compared to 
the Hubble expansion rate of the Universe $H(T)$ at temperature $T=M_i$ with
\be
\Gamma_{N_i} = \frac{(Y^\dagger Y)_{ii}M_i}{8\pi},\;\;\;\;\;\;\;\;
H(T)=\frac{2}{3}\sqrt{\frac{g_{*}\pi^{3}}{5}}\frac{T^{2}}{M_{pl}},
\label{eq-08:decay_Hubble}
\ee
where $M_{pl}=1.22 \times 10^{19}$ GeV is the Planck mass, 
$g_*$ (=106.75 for the SM excluding RH neutrinos) 
is the total number of relativistic degrees of freedom
contributing to the energy density of the Universe.

To quantify the departure from thermal equilibrium, we define the \emph{decay parameter}
as follows
\be
K_i \equiv \frac{\Gamma_{N_i}}{H(M_i)} = \frac{\widetilde m_i}{m_*},
\label{eq-08:decay_para}
\ee
where $\widetilde m_i$ is the \emph{effective neutrino mass} defined as\cite{Plumacher:1996kc}
\be
\widetilde m_i \equiv \frac{(Y^\dagger Y)_{ii} v^2 }{M_i},
\label{eq-08:eff_mass}
\ee
with $m_* \equiv \frac{16\pi^2 v^2}{3 M_{pl}}\sqrt{\frac{g_* \pi}{5}}
\simeq 1 \times 10^{-3}$ eV. The regimes where $K_i \ll 1$, $K_i \approx 1$ and $K_i \gg 1$
are respectively known as weak, intermediate, and strong washout regimes.

\subsection{CP asymmetry}
\label{sec-08:CPasy}

\begin{figure}
\begin{center}
\includegraphics[scale=0.6]{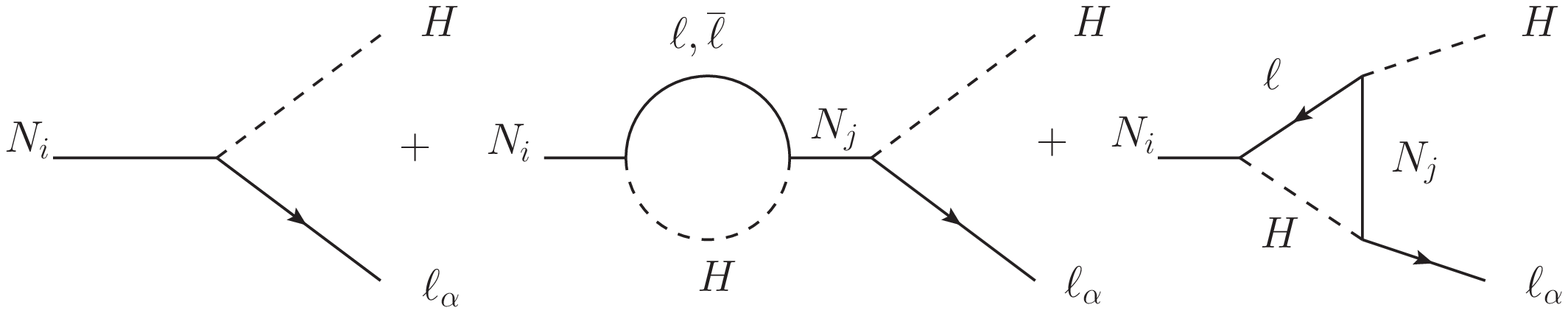}
\end{center}
\caption{The CP asymmetry in type-I seesaw leptogenesis results
from the interference between tree and 1-loop wave 
and vertex diagrams. For the 1-loop wave diagram, there
is an additional contribution from $L$-conserving diagram
to the CP asymmetry which vanishes when summing 
over lepton flavours.}
\label{fig-08:typeI_CP}
\end{figure}

The CP asymmetry in the decays of RH neutrinos $N_i$ can be defined as
\be
\epsilon_{i\alpha} = \frac{\gamma\left(N_i \to \ell_\alpha H\right)
-\gamma\left(N_i \to \overline{\ell_\alpha} H^*\right)}
{\sum_\alpha \gamma\left(N_i \to \ell_\alpha H\right)
+\gamma\left(N_i \to \overline{\ell_\alpha} H^*\right)}
\equiv \frac{\Delta \gamma_{N_i}^\alpha}{\gamma_{N_i}},
\label{eq-08:typeI_CP_def}
\ee
where $\gamma(i \to f)$ is the thermally averaged decay rate defined 
as\footnote{Here the Pauli-blocking and Bose-enhancement statistical factors have been ignored 
and we also assume Maxwell-Boltzmann distribution for the particle $i$ i.e. $f_i = e^{-E_i/T}$.
See Refs.~\cite{Covi:1997dr,Giudice:2003jh} for detailed studies of their effects.}
\bea
\gamma(i \to f) \equiv \int \frac{d^3p_i}{(2\pi)^3 2E_i}\frac{d^3p_f}{(2\pi)^3 2E_f}
(2\pi)^4\delta^{(4)}(p_i - p_f)|\mathcal{A}(i \to f)|^2 e^{-E_i/T},
\label{eq-08:therm_rate}
\eea
where $\mathcal{A}(i \to f)$ is the decay amplitude. 
Ignoring all thermal effects~\cite{Covi:1997dr,Giudice:2003jh}, 
\eqn{eq-08:typeI_CP_def} simplifies to
\be
\epsilon_{i\alpha} = \frac{|\mathcal{A}_0\left(N_i \to \ell_\alpha H\right)|^2
-|\mathcal{A}_0\left(N_i \to \overline{\ell_\alpha} H^*\right)|^2}
{\sum_\alpha |\mathcal{A}_0\left(N_i \to \ell_\alpha H\right)|^2
+|\mathcal{A}_0\left(N_i \to \overline{\ell_\alpha} H^*\right)|^2},
\label{eq-08:CP_asym2}
\ee
where $\mathcal{A}_0(i \to f)$ denotes the decay amplitude at zero temperature. 
\Eqn{eq-08:CP_asym2} vanishes at tree level but is induced at 1-loop level through the interference
between tree and 1-loop diagrams shown in Figure \ref{fig-08:typeI_CP}.
There are two types of contributions from the 1-loop diagrams:
the self-energy or wave diagram (middle)~\cite{Flanz:1994yx} 
and the vertex diagram (right)~\cite{Fukugita:1986hr}. 
At leading order, we obtain the CP asymmetry~\cite{Covi:1996wh}:
\bea
\epsilon_{i\alpha}&=&\frac{1}{8\pi}\frac{1}{(Y^\dagger Y)_{ii}} 
\sum_{j\neq i} {\rm Im}\left[(Y^\dagger Y)_{ji} Y_{\alpha i} Y_{\alpha j}^*\right]
g\left(\frac{M_j^2}{M_i^2}\right)\nonumber \\ 
&& +\frac{1}{8\pi}\frac{1}{(Y^\dagger Y)_{ii}}
\sum_{j\neq i} {\rm Im}\left[(Y^\dagger Y)_{ij}Y_{\alpha i} Y_{\alpha j}^*\right]
\frac{M_i^2}{M_i^2-M_j^2},
\label{eq-08:CP_asym}
\eea
where the loop function is 
\be
g(x) = \sqrt{x}\left[\frac{1}{1-x}+1
-(1+x)\ln\left(\frac{1+x}{x}\right)\right].
\ee
The first term in \eqn{eq-08:CP_asym} comes from  $L$-violating
wave and vertex diagrams, while the second term is from the
$L$-conserving wave diagram.  The terms of the form
$(M_i^2-M_j^2)^{-1}$ in \eqn{eq-08:CP_asym} are from the wave diagram
contributions which can resonantly enhance the CP asymmetry if $M_i
\approx M_j$ (resonant leptogenesis scenario, see Section
\ref{sec-08:resonant})\footnote{Notice that the resonant term becomes
  singular in the degenerate limit $M_i = M_j$.  This singularity can
  be regulated by using for example an effective field-theoretical
  approach based on resummation~\cite{Pilaftsis:1997jf}.}.  Let us  also
note that at least two RH neutrinos are needed, otherwise the CP
asymmetry vanishes because the Yukawa couplings combination becomes real.

In the one flavour regime, we sum over the flavour index $\alpha$ 
in \eqn{eq-08:CP_asym} and obtain
\be
\epsilon_{i}\equiv\sum_\alpha \epsilon_{i\alpha} = \frac{1}{8\pi}\frac{1}{(Y^\dagger Y)_{ii}} 
\sum_{j\neq i} {\rm Im}\left[(Y^\dagger Y)_{ji}^2\right]
g\left(\frac{M_j^2}{M_i^2}\right),
\label{eq-08:CP_asym_sum}
\ee
where the second term in \eqn{eq-08:CP_asym} vanishes 
because the combination of the Yukawa couplings is real.

\subsection{Classical Boltzmann equations}
\label{sec-08:CBE}
We work in the one flavour regime and consider only the decays and
inverse decays of $N_1$.  If leptogenesis occurs at $T \gtrsim
10^{12}\,$GeV, then the charged lepton Yukawa interactions are out of
equilibrium, and this defines the one flavour regime.  The assumption
that only the dynamics of $N_1$ is relevant can be realized if for
example the reheating temperature after inflation is $ T_{RH} \ll
M_{2,3}$ such that $N_{2,3}$ are not produced.
In order to scale out the effect of the expansion of the Universe, 
we will introduce the {\it abundances},  i.e. the ratios  
of the particle densities $n_i=\int d^3p f_i$ to the entropy density 
$s=\frac{2\pi^2}{45} g_* T^3$: 
\be
Y_i \equiv \frac{n_i}{s}.
\label{eq-08:Y}
\ee
The evolution of the $N_1$ density and the lepton asymmetry 
$Y_{\Delta L} = 2 Y_{\Delta \ell} \equiv 2(Y_{\ell} - Y_{\bar\ell})$ \footnote{The factor of 2 
comes from the $SU(2)_L$ degrees of freedoms.}
can be described by the following classical Boltzmann equations (BE)\cite{Kolb:1980}
\bea
\frac{dY_{N_1}}{dz} &=& -D_1 (Y_{N_1} - Y_{N_1}^{eq}), \label{eq-08:BE_N} \\
\frac{dY_{\Delta L}}{dz} &=& \epsilon_1 D_1 (Y_{N_1} - Y_{N_1}^{eq}) - W_1 Y_{\Delta L},
\label{eq-08:BE_L}
\eea where $z \equiv M_1/T$ and the decay and washout terms are
respectively given by \bea D_1(z) &=& \frac{\gamma_{N_1}z}{sH(M_1)} =
K_1 z \frac{{\cal K}_1(z)}{{\cal K}_2(z)},\;\;\;\;\;\;\;\; W_1(z) =
\frac{1}{2}D_1(z) \frac{Y_{N_1}^{eq}(z)}{Y_{\ell}^{eq}}, \eea with
${\cal K}_n$ the n-th order modified Bessel function of second kind.
$Y_{N_1}^{eq}$ and $Y_\ell^{eq}$ read:\footnote{To write down a simple
  analytic expression for the equilibrium density of $N_1$, we assume
  Maxwell-Boltzmann distribution. However,
  follwing~\cite{Fong:2011yx}, the normalization factor $Y_\ell^{eq}$
  is obtained from a Fermi-Dirac distribution.}

\be
Y_N^{eq}(z) = \frac{45}{2\pi^4 g_*} z^2 {\cal K}_2 (z),\;\;\;\;\;\;\; 
Y_\ell^{eq} = \frac{15}{4\pi^2 g_*}.
\label{eq-08:eq_den}
\ee

From \eqn{eq-08:BE_N} and \eqn{eq-08:BE_L}, the solution for
$Y_{\Delta L}$ can be written down as follows 
\bea Y_{\Delta L}(z) & =
& Y_{\Delta L}(z_i) e^{-\int_{z_i}^{z} dz' W_1(z')}
- \int_{z_i}^{z} dz' 
\epsilon_1(z') \frac{dY_{N_1}}{dz'} e^{-\int_{z'}^{z} dz'' W_1(z'')}
\label{eq-08:YLsol}
\eea 
where $z_i$ is some initial temperature when $N_1$ leptogenesis
begins, and we have assumed that any preexisting lepton asymmetry
vanishes $Y^0_{\Delta L}(z_i)=0$.  Notice that ignoring thermal effects,
the CP asymmetry is independent of the temperature $\epsilon_1(z) =
\epsilon_1$ (c.f. \eqn{eq-08:CP_asym_sum}).

\subsubsection{Weak washout regime}
In the weak washout regime ($K_1 \ll 1$), the initial condition on the
$N_1$ density $Y_{N_1}(z_i)$ is important.
%
If we assume thermal initial abundance of $N_1$
i.e. $Y_{N_1}(z_i)=Y_{N_1}^{eq}(0)$, we can ignore the washout when
$N_1$ starts decaying at $z \gg 1$ and we have 
\be Y_{\Delta
  L}^{t}(\infty) \simeq -\epsilon_1\int_{0}^{\infty} dz'
\frac{dY_{N_1}^{eq}}{dz'} = \epsilon_1 Y_{N_1}^{eq}(0).
\label{eq-08:YL_therm}
\ee
On the other hand, if we have zero initial $N_1$ abundance  
i.e. $Y_{N_1}(z_i)=0$, we have to consider the opposite sign contributions 
to lepton asymmetry from the inverse decays when $N_1$ is being populated
($Y_{N_1} < Y_{N_1}^{eq}$) and from the period when $N_1$ starts decaying ($Y_{N_1} > Y_{N_1}^{eq}$).
%
Taking this into account the term which survives the partial
cancellations are given by~\cite{Buchmuller:2004nz} \footnote{This
  differs from the efficiency in Ref.~\cite{Buchmuller:2004nz} by the
  factor $\frac{12}{\pi^2}$, which is due to the different
  normalization $Y_\ell^{eq}$~\eqn{eq-08:eq_den}.}  \be Y_{\Delta
  L}^{0}(\infty) \simeq \frac{27}{16} \epsilon_1 K_1^2
\,Y_{N_1}^{eq}(0).
\label{eq-08:YL_ini}
\ee


\subsubsection{Strong washout regime}

In the strong washout regime ($K_1 \gg 1$) any lepton asymmetry generated 
during the $N_1$ creation phase is efficiently washed out. Here we adopt 
the \emph{strong washout balance
  approximation}\cite{Fong:2010up} which states that in the strong
washout regime, the lepton asymmetry at each instant takes the value
that enforces a balance between the production and the destruction
rates of the asymmetry.  Equating the decay and washout terms in
\eqn{eq-08:BE_L}, we have \be Y_{\Delta L}(z) \approx
-\frac{1}{W(z)}\epsilon_1\frac{dY_{N_1}}{dz} \simeq
-\frac{1}{W(z)}\epsilon_1\frac{dY_{N_1}^{eq}}{dz} = \frac{2}{z K_1}
\epsilon_1 Y_\ell^{eq}, \ee where in the second approximation, we
assume $Y_{N_1} \simeq Y_{N_1}^{eq}$.  The approximation no longer
holds when $Y_{\Delta L}$ freezes and this happens when the washout
decouples at $z_f$ i.e. $W(z_f) < 1$. Hence, the final lepton
asymmetry is given by\footnote{Compare this to a more precise
  analytical approximation in Ref.~\cite{Buchmuller:2004nz}.}  \be
Y_{\Delta L}(\infty)=\frac{2}{z_f K_1} \epsilon_1 Y_\ell^{eq} =
\frac{\pi^2}{6 z_f K_1} \epsilon_1 Y_{N_1}^{eq}(0).
\label{eq-08:YL_strong}
\ee 
The freeze out temperature $z_f$ depends mildly on $K_1$. For $K_1
= 10\,$-\,100 we have for example $z_f = 7\,$-\,10.  We also see that
independently of initial conditions, in the strong regime $Y_{\Delta
  L}(\infty)$ goes as $K_1^{-1}$.

\subsection{Baryon asymmetry from EW sphaleron}
\label{sec-08:LtoB}

The final lepton asymmetry $Y_{\Delta L}(\infty)$ can be conveniently
parametrized as follows 
\be Y_{\Delta L}(\infty) = \epsilon_1\eta_1
Y_{N_1}^{eq}(0),
\label{eq-08:YLpara}
\ee
where $\eta_1$ is known as the \emph{efficiency factor}. 
In the weak washout regime ($K_1 \ll 1$) 
from  \eqn{eq-08:YL_therm} 
we have
$\eta_1 = 1\,(=\frac{27}{16}K_1^2<1)$ for thermal (zero) initial $N_1$ abundance. 
In the strong washout regime ($K_1 \gg 1$), from \eqn{eq-08:YL_strong},
we have $\eta_1 = \frac{\pi^2}{6z_f K_1} < 1$.

If leptogenesis ends before EW sphaleron processes become active ($T \gtrsim 10^{12}$ GeV),
the $B-L$ asymmetry $Y_{\Delta_{B-L}}$ is simply given by
\be
Y_{\Delta_{B-L}} = -Y_{\Delta L}\,.
\ee
At the later stage, the $B-L$ asymmetry is partially transfered to a 
$B$ asymmetry by the EW sphaleron processes 
through the relation~\cite{Harvey:1990qw}
\be
Y_{\Delta B}(\infty) = \frac{28}{79}Y_{\Delta_{B-L}}(\infty)\,,
\label{eq-08:YBLtoYB}
\ee
that holds if sphalerons decouple before EWPT.  This
relation will change if the EW sphaleron processes decouple after the
EWPT~\cite{Harvey:1990qw,Inui:1993wv} or if threshold
effects for heavy particles like the top quark and Higgs are taken
into account~\cite{Inui:1993wv,Chung:2008gv}.

\subsection{Davidson-Ibarra bound}
\label{sec-08:DI}

Assuming a hierarchical spectrum of the RH neutrinos ($M_1 \ll M_2,\,
M_3$) and that the dominant lepton asymmetry is from the $N_1$ decays,
from \eqn{eq-08:CP_asym_sum} the CP asymmetry from $N_1$ decays can 
be written as 
\be
\epsilon_{1} =
-\frac{3}{16\pi}\frac{1}{(Y^\dagger Y)_{11}} 
\sum_{j\neq 1} {\rm Im}\left[(Y^\dagger Y)^2_{j1}\right]
\frac{M_1}{M_j}.
\label{eq-08:epsilon_1}
\ee
Assuming three generations of RH neutrinos ($n=3$) and using the 
Casas-Ibarra parametrization~\cite{Casas:2001sr} for
the Yukawa couplings
\be
Y_{\alpha i}=\frac{1}{v}
\left(\sqrt{D_{m_N}}R\sqrt{D_{m_\nu}}U_\nu^\dagger\right)_{\alpha i},
\label{eq-08:yukawa_CI}
\ee
where $D_{m_N}={\rm diag}(M_1,M_2,M_3)$, 
$D_{m_\nu}={\rm diag}(m_{\nu_1},m_{\nu_2},m_{\nu_3})$ 
and $R$ any complex orthogonal matrix satisfying $R^TR=RR^T=1$, \eqn{eq-08:epsilon_1} becomes
\be
\epsilon_{1} =
-\frac{3}{16\pi}\frac{M_1}{v^2}\frac{{\displaystyle \sum_i} m_{\nu_i} 
{\rm Im}(R^2_{1i})}
{{\displaystyle \sum_i} m_{\nu_i} |R_{1i}|^2}.
\ee
Using the orthogonality condition ${\displaystyle \sum_i} R^2_{1i}=1$,
we then obtain the Davidson-Ibarra (DI) bound~\cite{Davidson:2002qv}
\be
|\epsilon_{1}| \leq \epsilon^{DI} =
\frac{3}{16\pi}\frac{M_1}{v^2}(m_{\nu_3}-m_{\nu_1})
= \frac{3}{16\pi}\frac{M_1}{v^2}\frac{\Delta m_{atm}^2}{m_{\nu_1}+m_{\nu_3}},
\label{eq-08:DI_bound}
\ee
where $m_{\nu_3}$ ($m_{\nu_1}$) is the heaviest 
(lightest) light neutrino mass. Applying the DI bound
on eqs.~(\ref{eq-08:YLpara})--(\ref{eq-08:YBLtoYB}), 
and requiring that $Y_{\Delta B}(\infty) \geq Y_{B}^{CMB} \simeq 10^{-10}$, 
we obtain
\be
M_1 \left(\frac{0.1\,{\rm eV}}{m_{\nu_1}+m_{\nu_3}}\right) \,\eta_1^{max}(M_1) \gtrsim 
10^9 \,{\rm GeV},
\label{eq-08:leptobound}
\ee
where the $\eta_1^{max}(M_1)$ is the efficiency factor maximized with respect to $K_1$ \eqn{eq-08:decay_para}
for a particular value of $M_1$.
This allows us to make a plot of region which satisfies
\eqn{eq-08:leptobound} on the $(M_1,m_{\nu_1})$ plane and hence obtain
bounds on $M_1$ and $m_{\nu_1}$.  Many careful numerical studies have
been carried out and it was found that successful leptogenesis with a
hierarchical spectrum of the RH neutrinos requires $M_1 \gtrsim
10^9$~GeV~\cite{Davidson:2002qv,Buchmuller:2002rq,Ellis:2002xg} and
$m_{\nu_1} \lesssim
0.1$~eV~\cite{Buchmuller:2002jk,Buchmuller:2003gz,Buchmuller:2004nz,Nardi:2011zz}.
This bound implies that the RH neutrinos must be produced at
temperatures $T \gtrsim 10^9\,{\rm GeV}$ which in turn implies the
reheating temperature after inflation has to be $T_{RH} \gtrsim
10^{9}\,{\rm GeV}$ in order to have sufficient RH neutrinos in the
thermal bath.  To conclude this section, let us note that the DI bound
\eqn{eq-08:DI_bound}
holds if and only if all the following conditions apply:  \\
(1) $N_1$ dominates the contribution to leptogenesis. \\
(2) The mass spectrum of RH neutrinos are hierarchical $M_1 \ll M_2, M_3$. \\
(3) Leptogenesis occurs in the unflavoured regime $T \gtrsim 10^{12}$ GeV. \\
As we will see in the following sections, violation of one or more of
the above conditions allows us to lower somewhat the scale of
leptogenesis.

\vspace{1cm} 
\section{Lepton Flavour Effects}
\label{sec-08:flavour}

\subsection{When are lepton flavour effects relevant?}
\label{sec-08:fla_temp}

The first leptogenesis calculations were performed in the single
lepton flavour regime. In short, this amounts 
to assuming that the leptons and antileptons which couple to the 
lightest RH neutrino $N_1$ maintain their coherence as flavour 
superpositions throughout the leptogenesis 
era, that is  $\ell_1 = \sum_\alpha c_{\alpha 1} \ell_\alpha$ 
and $\overline\ell_1'=\sum_\alpha c^{'*}_{\alpha 1} \overline\ell_\alpha$.  
Note that at the tree-level the coefficients $c$ and $c'^*$ are simply the Yukawa
couplings: $c_{\alpha 1} = Y_{\alpha 1}$ and $c_{\alpha 1}'^* =
Y^*_{\alpha 1}$. However it should be kept in mind that since CP is
violated by loops, beyond the tree level approximation the antilepton
state $\overline{\ell_1'}$ is not the CP conjugate of the $\ell_1$,
that is $c_{\alpha 1}' \neq c_{\alpha 1}$. 

The single flavour regime is realized only at very high temperatures
($T \gtrsim 10^{12}$ GeV) when both $\ell_1$ and $\ell_1'$ remain
coherent flavour superpositions, and thus are the correct states to
describe the dynamics of leptogenesis.  However, at lower temperatures
scatterings induced by the charged lepton Yukawa couplings occur at a
sufficiently fast pace to distinguish the different lepton flavours,
$\ell_1$ and $\ell_1'$ decohere in their flavour components, and the
dynamics of leptogenesis must then be described in terms of the
flavour eigenstates $\ell_\alpha$. Of course, there is great interest
to extend the validity of quantitative leptogenesis studies also at
lower scale $T \lesssim 10^{12}$ GeV, and this requires accounting for
flavour effects.
The role of lepton flavour in leptogenesis was first discussed in
Ref.~\cite{Barbieri:1999ma}, however the authors did not highlight in
what the results were significantly different from the single flavour
approximation. Therefore, until the importance of flavour effects was
fully clarified in Refs.~\cite{Abada:2006fw,Nardi:2006fx,Abada:2006ea}
they had been included in leptogenesis studies only in a few
cases~\cite{Endoh:2003mz,Abada:2004wn,
  Vives:2005ra,Fujihara:2005pv,Pilaftsis:2005rv}.
Nowadays lepton flavour effects have been investigated in full
detail~\cite{Branco:2006ce,Pascoli:2006ie,Pascoli:2006ci,Antusch:2006cw,
  Antusch:2006gy,Antusch:2007km,Blanchet:2006be,Blanchet:2006ch,
  DeSimone:2006dd,DeSimone:2007rw,DeSimone:2007pa,Cirigliano:2007hb,Cirigliano:2009yt,Fong:2010qh}
and are a mandatory ingredient of any reliable analysis of
leptogenesis.

The specific temperature when leptogenesis becomes
sensitive to lepton flavour dynamics can be estimated by requiring
that the rates of processes $\Gamma_\alpha$ ($\alpha=e,\mu,\tau$) that
are induced by the charged lepton Yukawa couplings $h_\alpha$ become
faster than the Universe expansion rate $H(T)$. An approximate
relation gives~\cite{Campbell:1992jd,Cline:1993bd}
\be
\Gamma_\alpha(T) \simeq 10^{-2} h_\alpha^2 T\,,
\label{eq-08:fla_int_rate}
\ee
which implies that\footnote{In supersymmetric case, since $h_\alpha = m_\alpha/(v_u \cos\beta)$,
we have $T \lesssim T_\alpha \, (1+\tan^2\beta)$.}
\be
\Gamma_\alpha(T) > H(T) \qquad {\rm when} \qquad  
T \lesssim  T_\alpha \,, 
\label{eq-08:fla_eq_rate}
\ee
where $T_e \simeq 4\times 10^4\,$GeV, $T_\mu \simeq 2\times 10^9
\,$GeV, and $T_\tau \simeq 5\times 10^{11}\,$GeV. Notice that to fully
distinguish the three flavours it is sufficient that the $\tau$ and
$\mu$ Yukawa reactions attain thermal equilibrium. It has been pointed
out that besides being faster than the expansion of the Universe, the
charged lepton Yukawa interactions should also be faster than the
$N_1$
interactions~\cite{Nardi:2006fx,Blanchet:2006ch,DeSimone:2006dd}.  In
general whenever $\Gamma_\tau (M_1) > H(M_1)$ we also have
$\Gamma_\tau (M_1) > \Gamma_{N_1} (M_1)$.  However, there exists
parameter space where $\Gamma_\tau (M_1) > H(M_1)$ but $\Gamma_\tau
(M_1) < \Gamma_{N_1} (M_1)$. This scenario was studied in
Ref.~\cite{Blanchet:2006ch}.

\subsection{The effects on CP asymmetry and washout}

The CP violation in $N_i$ decays can manifest itself in two ways~\cite{Nardi:2006fx}:\\ [3pt]
(i) The leptons and antileptons are produced at different rates, \be
\gamma_{i} \neq \bar\gamma_{i},
\label{eq-08:type_i}
\ee
where $\gamma_{i} \equiv \gamma(N_i \to \ell_i H)$ 
and $\bar\gamma_{i} \equiv \gamma(N_i \to \overline{\ell_i'} H^*)$. \\ [3pt]
\noindent (ii) The leptons and antileptons produced are not CP conjugate states,
\be
CP(\overline{\ell_i'}) = \ell_i' \neq \ell_i,
\label{eq-08:type_ii}
\ee
that is, due to loops effects they are slightly misaligned in flavour space.

We can rewrite the CP asymmetry for $N_i$ decays from \eqn{eq-08:typeI_CP_def} as follows
\be
\epsilon_{i\alpha} 
 =  \frac{P_{i\alpha} \gamma_i - \bar P_{i\alpha} \bar\gamma_i}
{\gamma_i + \bar\gamma_i} 
 =  \frac{P_{i\alpha} + \bar P_{i\alpha}}{2}\, \epsilon_i
+ \frac{P_{i\alpha} - \bar P_{i\alpha}}{2} 
 \simeq P_{i\alpha}^0 \epsilon_i + \frac{\Delta P_{i\alpha}}{2},
\label{eq-08:fla_CP}
\ee
where terms of order ${\cal O}(\epsilon_i\,\Delta P_{i\alpha})$ and higher have been neglected.
$P_{i\alpha}$ is the projector from state $\ell_i$ into flavour state $\ell_\alpha$
and $\Delta P_{i\alpha} = P_{i\alpha} - \bar P_{i\alpha}$. 
At tree level, clearly, $P_{i\alpha} = \bar P_{i\alpha} \equiv P_{i\alpha}^0$ where 
the tree level flavour projector is given by
\be
P_{i\alpha}^0 = \frac{Y_{\alpha i} Y_{\alpha i}^*}{(Y^\dagger Y)_{ii}}.
\label{eq-08:fla_proj}
\ee From \eqn{eq-08:fla_CP}, we can identify the two types of CP
violation, the first term being of type (i) \eqn{eq-08:type_i} while
the second being of type (ii) \eqn{eq-08:type_ii}.  Since $\sum_\alpha
P_{i\alpha} = \sum_\alpha \bar P_{i\alpha} = 1$, when summing over
flavour indices $\alpha$, the second term vanishes $\sum_\alpha \Delta
P_{i\alpha} = 0$. Note that the lepton-flavour-violating but
$L$-conserving terms in the second line of \eqn{eq-08:CP_asym} is part
of type (ii).  In fact, they come from $d=6$ $L$-conserving operators
which have nothing to do with the unique $d=5$ $L$-violating operator
(the Weinberg operator~\cite{Weinberg:1979sa}) responsible for
neutrino masses. However, in some cases they can still dominate the CP
asymmetries but, as we will see in Section \ref{sec-08:LFE}, lepton
flavour equilibration effects~\cite{AristizabalSierra:2009mq} then
impose important constraints on their overall effects.  Note also that
due to flavour misalignment, the CP asymmetry in a particular flavour
direction $\epsilon_{i\alpha}$ can be much larger and even of opposite
sign from the total CP asymmetry $\epsilon_i$.  In fact the relevance
of CP violation of type (ii) in the flavour regimes is what allows to
evade the DI bound \eqn{eq-08:DI_bound}. As regards the washout of the
lepton asymmetry of flavour $\alpha$, it is proportional to
\be 
W_{i\alpha} \propto P_{i\alpha}\gamma_i + \bar P_{i\alpha} \bar\gamma_i 
\simeq P_{i\alpha}^0 W_i,
\label{eq-08:fla_washout}
\ee
which results in a reduction of washout by a factor of $P_{i\alpha}^0\leq 1$
compared to unflavoured case.
As we will see next, the new CP-violating sources from flavour effects
and the reduction in the washout could result in great enhancement of
the final lepton asymmetry and,  as was first pointed out in
Ref.~\cite{Nardi:2006fx}, leptogenesis with a vanishing total CP
asymmetry $\epsilon_{i} = 0$ also becomes possible.

\subsection{Classical flavoured Boltzmann equations}

Here again we only consider leptogenesis from the decays and inverse
decays of $N_1$. In this approximation, the BE for $Y_{N_1}$ is still
given by \eqn{eq-08:BE_N} while the BE for $Y_{\Delta L_\alpha}$ the
lepton asymmetry in the flavour $\alpha$ is given by\footnote{To study
  the transition between different flavour regimes (from one to two or
  from two to three flavours), a density matrix formalism has to be
  used~\cite{Abada:2006fw,DeSimone:2006dd,Beneke:2010dz}.}  \bea
\frac{dY_{\Delta L_\alpha}}{dz} &=& \epsilon_{1\alpha} D_1 (Y_{N_1} -
Y_{N_1}^{eq}) - P_{1\alpha}^0 W_1 Y_{\Delta L_\alpha}.
\label{eq-08:BE_L_fla}
\eea
Notice that as long as $L$ violation from sphalerons is neglected (see
section \ref{sec-08:improving}) the BE for $Y_{\Delta L_\alpha}$ are
independent of each other, and hence the solutions  for
the weak and strong washout regimes are given respectively by
\eqn{eq-08:YL_ini}) and \eqn{eq-08:YL_strong}, after replacing
$\epsilon_1 \to \epsilon_{1\alpha}$ and $K_1 \to
K_{1\alpha} \equiv P_{1\alpha}^0 K_1$.

As an example let us assume that leptogenesis occurs around $T \sim
10^{10}\,$GeV, that is in the two-flavour regime. Due to the fast $\tau$ Yukawa interactions 
$\ell_1 (\ell_1')$ gets projected onto $\ell_\tau (\ell_\tau')$ and a
coherent mixture of $e+\mu$ eigenstate $\ell_{e+\mu} (\ell_{e+\mu}')$.
For illustrative purpose,
here we  consider a scenario in which  lepton flavour effects
are most prominent. We take 
both $K_{1\tau}, K_{1e+\mu} \gg 1$, so that both $Y_{\Delta L_\tau}$ and
$Y_{\Delta L_{e+\mu}}$ are in the strong regime. From
\eqn{eq-08:YL_strong} we can write down the solution:
\bea
Y_{\Delta L}(\infty) & = & Y_{\Delta L_\tau}(\infty) + Y_{\Delta L_{e+\mu}}(\infty) \nonumber \\
             & = & \frac{\pi^2}{6z_f K_1} Y_{N_1}^{eq}(0) 
        \left( \frac{\epsilon_{1\tau}}{P_{1\tau}^0} 
        + \frac{\epsilon_{1e+\mu}}{P_{1 e+\mu}^0} \right) \nonumber \\
    & \simeq & \frac{\pi^2}{3z_f K_1} \epsilon_1 Y_{N_1}^{eq}(0) +
       \frac{\pi^2}{12z_f K_1} Y_{N_1}^{eq}(0) \left( \frac{\Delta P_{1\tau}}{P_{1\tau}^0} 
   + \frac{\Delta P_{1e+\mu}}{P_{1 e+\mu}^0} \right),
\label{eq-08:YL_strong_2fla}
\eea 
where in the last line we have used \eqn{eq-08:fla_CP}.  If
$P_{1\tau}^0 \simeq P_{1e+\mu}^0$, then since $\Delta P_{1\tau} +
\Delta P_{1e+\mu} = 0$ the second term approximately cancels, and  
\eqn{eq-08:YL_strong_2fla} reduces to 
\be Y_{\Delta L}(\infty) \simeq \frac{\pi^2}{3z_f
  K_1}\epsilon_1 Y_{N_1}^{eq}(0)\,. 
\label{eq-08:YL_flaeq_strong}
\ee
We see that the final asymmetry is enhanced by a factor of 2 compared
to the unflavoured case.  If there exists some hierarchy between the
flavour projectors, then the second term in \eqn{eq-08:YL_strong_2fla}
plays an important role and can further enhance the  asymmetry.
For example we can have  $P_{1\tau}^0 >
P_{1e+\mu}^0$ while $\Delta P_{1\tau} \ll \Delta P_{1e+\mu}$. In this
case, the second term can dominate over the first term. 
Finally from \eqn{eq-08:YL_strong_2fla} we also notice that
leptogenesis with $\epsilon_1 = 0$, the so-called \emph{purely
  flavoured leptogenesis} (PFL)\footnote{This can also refer to the
  case where the total CP asymmetry is negligible $\epsilon_1 \approx
  0$.}, can indeed
proceed~\cite{Nardi:2006fx,AristizabalSierra:2009bh,GonzalezGarcia:2009qd,Antusch:2009gn,Racker:2012vw}.
In this scenario some symmetry has to be imposed to realize the
condition $\epsilon_1 = 0$, as for example an approximate global
lepton number $U(1)_L$. In the limit of exact $U(1)_L$ the active
neutrinos will be exactly massless. Instead of the seesaw mechanism,
the small neutrino masses is explained by $U(1)_L$ which is slightly
broken by a small parameter $\mu$ (the ``inverse
seesaw'')~\cite{Mohapatra:1986bd} which is technically natural since
the Lagrangian exhibits an enhanced symmetry when $\mu \to
0$~\cite{'tHooft:1979bh}.  In the next section, we will discuss
another aspect of flavour effects which are in particular crucial for
PFL.

\subsection{Lepton flavour equilibration}
\label{sec-08:LFE}

Another important effect is lepton flavour equilibration
(LFE)~\cite{AristizabalSierra:2009mq}. LFE processes violate lepton
flavour but conserve total lepton number e.g. $\ell_\alpha H \to
\ell_\beta H$, and can proceed e.g.  via off-shell exchange of
$N_{2,3}$. In thermal equilibrium, LFE processes can
quickly equilibrate the asymmetries generated in different
flavours. In practice this would be equivalent to a situation where
all the flavour projectors \eqn{eq-08:fla_proj} are equal, in which
case the flavoured BE \eqn{eq-08:BE_L_fla} can be summed up into a
single BE:
\bea 
\frac{dY_{\Delta L}}{dz} &=& \epsilon_{1} D_1
(Y_{N_1} - Y_{N_1}^{eq}) - P_{1\alpha}^0 W_1 Y_{\Delta L}, 
\eea
where $P_{1\alpha}^0 = 1/2\,(1/3)$ in the two (three) flavours
regime. In this case the BE is just like the unflavoured case but with
a reduced washout which, in the strong washout regime, would result in
enhancement of a factor of 2 (3) in the two (three) flavours regime
(c.f. \eqn{eq-08:YL_flaeq_strong}). 
Clearly, LFE can make PFL with $\epsilon_1 = 0$
impotent~\cite{AristizabalSierra:2009mq,Fong:2010up}.  Since LFE
$N_{2,3}$ processes scale as $T^3$ while the Universe expansion scales
as $T^2$, in spite of the fact that PFL evades the DI bound, they
eventually prevent the possibility of lowering too much the
leptogenesis scale.  A generic study in PFL scenario taking into
account LFE effects concluded that successful leptogenesis still
requires $M_1 \gtrsim 10^8$ GeV~\cite{Antusch:2009gn}.  A more
accurate study in the same direction recently carried out in
Ref.~\cite{Racker:2012vw}, showed that in fact the leptogenesis scale
can be lowered down to $M_1 \sim 10^6$ GeV.

\vspace{1cm}

\section{Beyond the Basic Boltzmann Equations}
\label{sec-08:improving}

Within factors of a few, the amount of baryon asymmetry that is
generated via leptogenesis in $N_1$ decays is determined essentially
by the size of the (flavoured) CP asymmetries and by the rates of the
(flavoured) washout reactions. However, to obtain more precise results
(say, within an ${\cal O}(1)$ uncertainty) several additional effects
must be taken into account, and the formalism must be extended well
beyond the basic BE discussed in the previous Sections. 
In the following we review some of the most important sources of
corrections, namely spectator processes (Section \ref{sec-08:spec}),
scatterings with top quarks and gauge bosons (Section
\ref{sec-08:scatterings}), thermal effects (Section
\ref{sec-08:thermal}), contributions from heavier RH neutrinos
(Section \ref{sec-08:heavier}), and we also discuss the role of
quantum corrections evaluated in the quantum BE approach (Section
\ref{sec-08:quantum}).  Throughout this review we use integrated BE,
i.e. we assume kinetic equilibrium for all particle species, and thus
we use particles densities instead than particles distribution
functions.  Corrections arising from using non-integrated BE have been
studied for example in
Refs.~\cite{Basboll:2006yx,Garayoa:2009my,HahnWoernle:2009qn,HahnWoernle:2009en},
and are generally subleading.

\subsection{Spectator processes}
\label{sec-08:spec}

Reactions that without involving violation of $B-L$ can still affect
the final amount of baryon asymmetry are classified as {\it
  ``spectator processes''}~\cite{Buchmuller:2001sr,Nardi:2005hs}.  The
basic way through which they act is that of redistributing the
asymmetry generated in the lepton doublets among the other particle
species. Since the density asymmetries of the lepton doublets are what
weights the rates of the washout processes, it can be expected that
spectator processes would render the washouts less effective and
increase the efficiency of leptogenesis.  However, in most cases this
is not true: proper inclusion of spectator processes implies
accounting for all the particle asymmetries, and in particular also for
the density asymmetry of the Higgs $Y_{\Delta H}$~\cite{Nardi:2005hs}.
This was omitted in Section~\ref{sec-08:basic} but in fact has to
be added to the density asymmetry of the leptons $Y_{\Delta \ell}$ in weighting for
example washouts from inverse decays.~\Eqn{eq-08:BE_L} would then
become:
\begin{equation}
\frac{dY_{\Delta L}}{dz} = \epsilon_1 D_1 (Y_{N_1} - Y_{N_1}^{eq}) 
- 2\,\left(Y_{\Delta \ell}+Y_{\Delta H}\right)  W_1 
\label{eq-08:BE_Lspect}
\end{equation}
where the factor of two in front of the washout term counts the
leptons and Higgs gauge multiplicity.  Clearly, in some regimes in
which $Y_{\Delta \ell}$ and $Y_{\Delta H}$ are not sufficiently
diluted by interacting with other particles, this can have the effect
of enhancing the washout rates and suppressing the efficiency.

In the study of spectator processes it is fundamental to specify the
range of temperature in which leptogenesis occurs. This is because at
each specific temperature $T$, particle reactions must be treated in a
different way depending if their characteristic time scale $\tau$
(given by inverse of their thermally averaged rates)
is~\cite{Fong:2010qh,Fong:2010bv}
\begin{itemize} \itemsep=-3pt
\item[(1)] much shorter than the age of the Universe:\quad $\tau \ll t_U(T)$; 
\item[(2)] much larger than the age of the Universe:\quad $\tau \gg t_U(T)$; 
\item[(3)] comparable with the  Universe age:\quad $\tau \sim t_U(T)$.
\end{itemize}
Spectator processes belong to the first type of reactions which occur
very frequently during one expansion time.  Their effects can be
accounted for by imposing on the thermodynamic system the chemical
equilibrium condition appropriate for each specific reaction, that is
$\sum_I \mu_I = \sum_F \mu_F$, where $\mu_I$ denotes the chemical
potential of an initial state particle, and $\mu_F$ that of a final
state particle\footnote{The relation between chemical potentials and
  particle density asymmetries is given in~\eqn{eq-08:Dnmu}.}.  The
numerical values of the parameters that are responsible for these
reactions only determine the precise temperature $T$ when chemical
equilibrium is attained but, apart from this, have no other relevance,
and do not appear explicitly in the formulation of the problem.
Reactions of type (2) cannot have any effect on the system, since they
basically do not occur. All physical processes are blind to the
corresponding parameters, that can be set to zero in the effective
Lagrangian.  In most cases this results in exact global symmetries
corresponding to conserved charges, and these conservation laws impose
constraints on the particle chemical potentials.  Reactions of type
(3) in general violate some symmetry, and thus spoil the corresponding
conservation conditions, but are not fast enough to enforce chemical
equilibrium.  These are the only reactions that need to be studied by
means of BE, and for which the precise value of the parameters that
control their rates is of utmost importance.

A simple case to illustrate how to include spectator processes is the
one flavour regime at particularly high temperatures (say $T\gsim
10^{13}\,$GeV).  The Universe expansion is fast implying that except
for processes induced by the large Yukawa coupling of the top and for
gauge interactions, all other $B-L$-conserving reactions fall in class
(ii). Then there are several conserved quantities as for example the
total number density asymmetries of the RH leptons as well as those 
of all the quarks except the top. Since electroweak sphalerons are also out of
equilibrium,  $B$ is conserved too (and vanishing, if we assume that
there is no preexisting asymmetry). $B=0$ then translates in the
condition:
\begin{equation}
  \label{eq-08:Bcons}
 2 Y_{\Delta Q_3}+Y_{\Delta t}=0 \,, 
\end{equation}
where $Y_{\Delta Q_3}$ is the density asymmetries of one degree of
freedom of the top  $SU(2)_L$ doublet and color triplet which,
being gauge interactions in equilibrium, is the same for all the six
gauge components, and $Y_{\Delta t}$ is the density asymmetry
of the $SU(2)_L$ singlet top.  Hypercharge is always conserved, yielding
\begin{equation}
  \label{eq-08:Ycons}
  Y_{\Delta Q_3}+ 2 Y_{\Delta t}  - Y_{\Delta \ell}+Y_{\Delta H}=0\,.  
\end{equation}
Finally, in terms of density asymmetries chemical equilibrium for the
top-Yukawa related reactions $\mu_{Q_3}+\mu_H=\mu_t$ translates into
\begin{equation}
  \label{eq-08:topYuk}
  Y_{\Delta Q_3}+\frac{1}{2}Y_{\Delta H}= Y_{\Delta t}\,. 
\end{equation}
We have three conditions for four density asymmetries, which allows
to express the Higgs density asymmetry in terms of the density
asymmetry of the leptons as $Y_{\Delta H}=\frac{2}{3} Y_{\Delta
  \ell}$. Moreover, given that only the LH lepton degrees of freedom
are populated, we have $Y_{\Delta L}= 2 Y_{\Delta\ell}$ so that the
coefficient weighting $W_1$ in \eqn{eq-08:BE_Lspect} becomes $
2\,\left(Y_{\Delta \ell}+Y_{\Delta H}\right)=\frac{5}{3}Y_{\Delta L}$ 
and the washout is accordingly stronger.

With decreasing temperatures, more reactions attain chemical
equilibrium, and accounting for spectator processes becomes accordingly
more complicated.  When the temperature drops below $T\sim
10^{12}\,$GeV, EW sphalerons are in equilibrium, and baryon number is
no more conserved. Then the condition \eqn{eq-08:Bcons} is no more
satisfied and, more importantly, the BE \eqn{eq-08:BE_Lspect} is no
more valid since sphalerons violate also lepton number with
in-equilibrium rates. However, sphalerons conserve $B-L$, which is
then violated only by slow reactions of type (3), and we should then
write down a BE for this quantity. Better said, since at $T\lsim
10^{12}\,$GeV all the third generation Yukawa reactions, including the
ones of the $\tau$-lepton, are in equilibrium, the dynamical regime is
that of two flavours in which the relevant quasi-conserved charges are
$\Delta_\tau = B/3-L_\tau$ and $\Delta_{e\mu} = B/3-L_{e\mu}$.
The fact that only two charges are relevant is because 
 there is always a direction in $e$-$\mu$ space which remains
decoupled from $N_1$. The corresponding third charge
$\Delta_{e\mu}'$ is then exactly conserved, its value can be set to zero
and the corresponding BE dropped. In this regime, the BE corresponding
to~\eqn{eq-08:BE_Lspect}  becomes:
\begin{equation}
- \frac{dY_{\Delta_\alpha}}{dz} = \epsilon_\alpha D_1 (Y_{N_1} - Y_{N_1}^{eq}) 
- 2\,\left(Y_{\Delta \ell_\alpha}+Y_{\Delta H}\right)  W_1 \qquad\quad 
(\alpha = \tau,e\mu)\,.  
\label{eq-08:BE_Lspect2fl}
\end{equation}
To rewrite these equations in a solvable closed form, $Y_{\Delta
  \ell_\tau}\,$, $Y_{\Delta \ell_{e\mu}}\,$ and $Y_{\Delta H}$ must be
expressed in terms of the two charge densities $Y_{\Delta_\tau}$ and
$Y_{\Delta_{e\mu}}$. This can be done by imposing the hypercharge
conservation condition~\eqn{eq-08:Ycons} and the chemical equilibrium
conditions that, in addition to~\eqn{eq-08:topYuk}, are appropriate
for the temperature regime we are considering. They
are~\cite{Nardi:2005hs}: 1. QCD sphaleron equilibrium; 2. EW sphaleron
equilibrium; 3. $b$-quark and $\tau$-lepton Yukawa equilibrium. The
`rotation' from the particle density asymmetries $Y_{\Delta
  \ell_\alpha}\,, Y_{\Delta H}$ to the charge densities
$Y_{\Delta_{\alpha}}$ can be expressed in terms of the $A$ matrix
introduced in~\cite{Barbieri:1999ma} $Y_{\Delta \ell_\alpha}=
A^\ell_{\alpha\beta} Y_{\Delta_\beta}$ $(\alpha,\beta=\tau,e\mu)$ and
$C$-vector $Y_{\Delta H}= C^H_{\alpha} Y_{\Delta_\alpha}$ introduced
in~\cite{Nardi:2006fx}. For the present case, with the ordering
$(e\mu,\tau)$ they are~\cite{Nardi:2006fx}:
\begin{equation} 
A^\ell = \frac{1}{460}\pmatrix{
196 &  -24 \cr
-9 &  156 }
\qquad \hbox{\rm and} \qquad 
C^H = \frac{1}{230}(41,\> 56)\,.  
\label{case4}
\end{equation}
It is important to stress that in each temperature regime there are
always enough constraints (conservation laws and chemical equilibrium
conditions) to allow to express all the relevant particle density
asymmetries in terms of the quasi-conserved charges
$Y_{\Delta_{\alpha}}$. This is because each time a conservation law
has to be dropped (like $B$ conservation above) it gets replaced by a
chemical equilibrium condition (like EW sphalerons equilibrium), and
each time the chemical potential of a new particle species becomes
relevant, it is precisely because a new reaction involving that
particle attains chemical equilibrium, enforcing the corresponding
condition.  As regards the quantitative corrections ascribable to
spectator processes, several numerical studies have confirmed that
they generally remain below order one. Thus, differently from flavour effects, 
for order of magnitude  estimates  they can be neglected.

\subsection{Scatterings and  CP violation in scatterings}
\label{sec-08:scatterings}

\begin{figure}
\begin{center}
\includegraphics[scale=0.8]{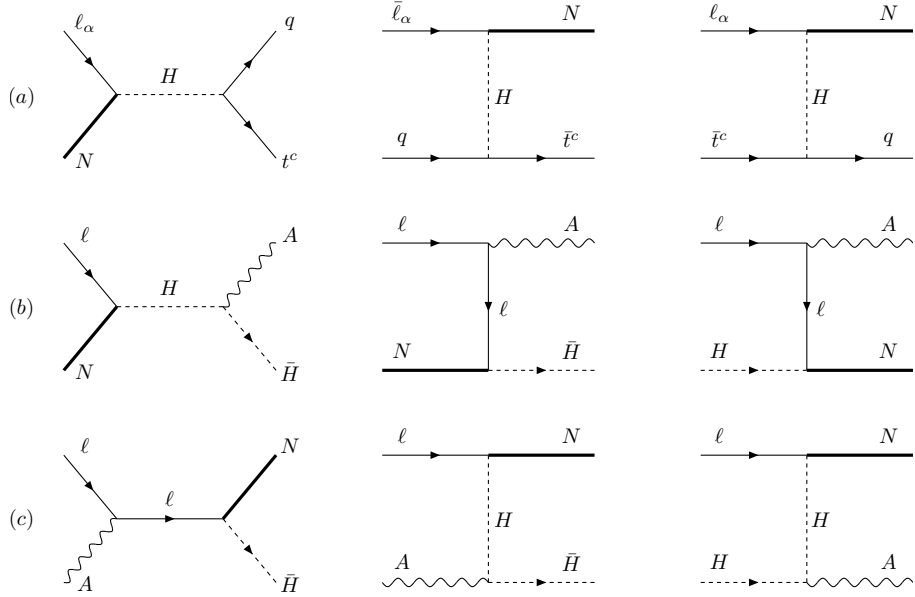}
\end{center}
\caption{Diagrams for  various $2\leftrightarrow 2$
  scattering processes:
 $(a)$ scatterings with
  the top quarks, $(b),\,(c)$ scatterings with the gauge bosons ($A=B,W_i$
  with $i=1,2,3$).
}
\label{fig-08:scatterings}
\end{figure}

Scattering processes are relevant for the production of the $N_1$
population, because decay and inverse decay rates are suppressed by a
time dilation factor $\propto M_1/T$. The $N_1 = \bar{N}_1$ particles
can be produced by scatterings with the top quark in $s$-channel
$H$-exchange $q t^c \to N \ell_\alpha$ and $\bar{q} \bar{t}^c \to N
\bar{\ell}_\alpha$, by $t$-channel $H$-exchange in $q\bar{\ell}_\alpha
\to N \bar{t}^c$, $\bar{q} \ell_\alpha \to N t^c$ and by $u$-channel
$H$-exchange in $\ell_\alpha\bar{t}^c \to N q$, $t^c \bar{\ell}_\alpha
\to N \bar{q} $, see the diagrams labeled (a) in
Figure \ref{fig-08:scatterings}.  Several scattering channels with gauge
bosons also contribute to the production of $N_1$. 
The corresponding diagrams are labeled $(b)$ and $(c)$ in the same figure.

When the effects of scatterings in populating the $N_1$ degree of
freedom are included, for consistency CP violation in scatterings must
also be included.  In doing so some care has to be put in treating
properly also all the processes of higher order in the couplings
($h_t^2 \lambda^4$, $g^2 \lambda^4$, where $g$ is a gauge coupling)
with an on-shell intermediate state $N_1$ subtracted out. This can be
done by following the procedure adopted in Ref.~\cite{Nardi:2007jp},
and we refer to that paper for details.

In first approximation, the CP asymmetry in scattering processes is
the same as in decays and inverse decays
\cite{Abada:2006ea,Davidson:2008bu}.  This result was first found in
Refs.~\cite{Pilaftsis:2003gt,Pilaftsis:2005rv,Anisimov:2005hr} for the
case of resonant leptogenesis, and was later derived in Ref.
\cite{Abada:2006ea} for the case of hierarchical $N_j$.  A full
calculation of the CP asymmetry in scatterings involving the top quark
was carried out in Ref.  \cite{Nardi:2007jp}, and the validity of
approximating it with the CP asymmetry in decays was analyzed, finding
that the approximation is generally good for sufficiently strong RH
neutrino hierarchies, e.g.  $M_2/M_1 \gg 10$.  Corrections up to
several tens of percent can appear around temperatures of order $T
\sim M_2/10$, and can be numerically relevant in case of milder
hierarchies.

Regarding the scattering processes with gauge bosons such as
$N\ell_\alpha \to A\bar H$ $NH \to A\bar\ell_\alpha$ or $NA \to
\ell_\alpha H$, their effects in leptogenesis were estimated in Ref.
\cite{Nardi:2007jp} under the assumption that it can also be
factorized in terms of the decay CP asymmetry.  However, with respect
to scatterings involving the top quark, there is a significant
difference that now box diagrams in which the gauge boson is attached
to a lepton or Higgs in the loop of the vertex-type diagrams are also
present, leading to more complicated expressions that were explicitly
calculated in Ref. \cite{Fong:2010bh}.  There it was shown that the
presence of box diagrams implies that for scatterings with gauge
bosons the CP asymmetry is different from the decay CP asymmetry even
for hierarchical RH neutrinos.  Still, this difference
remains within a factor of two \cite{Fong:2010bh} so that related
effects are in any case not very large.  In general, it turns out that
CP asymmetry in scatterings is more relevant at high temperatures ($T
> M_1$) when the scattering rates are larger than the decay
rate. Hence, it can be of some relevance to the final value of the
baryon asymmetry when some of the lepton flavours are weakly washed
out, and some memory of the asymmetries generated at high temperature
is preserved in the final result.

\subsection{Thermal corrections}
\label{sec-08:thermal}

%
At the high temperatures at which leptogenesis occurs, the light
particles involved in the leptogenesis processes are in equilibrium with the hot
plasma. Thermal effects give corrections to several ingredients in
the analysis: (i) coupling constants, (ii) particle propagators
(leptons, quarks, gauge bosons and the Higgs) and (iii) CP-violating
asymmetries, which we briefly discuss below.  A detailed study of
thermal corrections can be found in Ref. \cite{Giudice:2003jh}.

\subsubsection{Coupling constants}
\label{sec-08:couplings}
Renormalization of gauge and Yukawa couplings in a thermal plasma is
studied in Ref. \cite{Kajantie:1995dw}. In practice, it is a good
approximation to use the zero-temperature renormalization group
equations for the couplings, with a renormalization scale $\Lambda\sim
2\pi T$ \cite{Giudice:2003jh}. The value $\Lambda > T$ is related to
the fact that the average energy of the colliding particles in the
plasma is larger than the temperature.

The renormalization effects for the neutrino couplings are also well
known \cite{Casas:1999tg,Antusch:2003kp}. In the non-supersymmetric
case, to a good approximation these effects can be described by a
simple rescaling of the low energy neutrino mass matrix $m(\mu)=r\cdot
m$, where $1.2\lsim r\lsim 1.3$ for $10^8\,$GeV\,$\lsim \mu\lsim
10^{16}\,$GeV~\cite{Giudice:2003jh}, and therefore can be accounted
for by increasing the values of the neutrino mass parameters (for
example, $\tilde m$) as measured at low energy by $\approx 20\%-30\%$
(depending on the leptogenesis scale).  In the supersymmetric case one
expects a milder enhancement, but uncertainties related with the
precise value of the top-Yukawa coupling can be rather large (see
Figure 3 in Ref.~\cite{Giudice:2003jh}).

\subsubsection{Decays and scatterings}
\label{sec-08:thermaldecays}

In the thermal plasma, any particle with sizable couplings to the
background  acquires a thermal mass which is proportional to the
plasma temperature. Consequently, decay and scattering rates get
modified.  Particle thermal masses have been thoroughly studied in
both the SM and the supersymmetric SM~\cite{Comelli:1996vm,
Elmfors:1993re,Cline:1993bd,Weldon:1989ys,Weldon:1982bn,Klimov:1981ka}. 
The singlet neutrinos have no gauge
interactions, their Yukawa couplings are generally small and, during
the relevant era, their bare masses are of the order of the temperature
or larger. Consequently, to a good approximation, corrections to their
masses can be neglected. We thus need to account for the thermal
masses of the leptons and Higgs doublets and, when scatterings are
included, also of the third generation quarks and of the gauge bosons
(and of their superpartners in the supersymmetric case).  For a
qualitative discussion, it is enough to keep in mind that, within the
leptogenesis temperature range, we have $m_H(T)\gsim m_{q_3,t}(T)
\gg m_\ell(T)$. The most important effects are related to four classes
of leptogenesis processes:

(i) {\it Decays and inverse decays.} Since thermal corrections to the
Higgs mass are particularly large ($m_H(T) \approx 0.4\, T$),
decays and inverse decays become kinematically forbidden in the
temperature range in which $m_H(T)-m_\ell(T)
<M_{N_1}<m_H(T)+m_\ell(T)$. For lower temperatures, the usual
processes $N_1\leftrightarrow \ell H$ can occur.  For higher
temperatures, the Higgs is heavy enough that it can decay:
$H\leftrightarrow \ell N_1$. A rough estimate of the kinematically
forbidden region yields $2 \lsim T/M_1\lsim 5$. The important point
is that these corrections are effective only at $T>M_1$.  In the
parameter region $\tilde m > 10^{-3}\,$eV, that is favoured by the
measurements of the neutrino mass-squared differences, the $N_1$ number
density and its $L$-violating reactions attain thermal equilibrium at
$T \approx M_1$ and erase quite efficiently any memory of the specific
conditions at higher temperatures. Consequently, in the strong washout
regime, these  corrections have practically no effect on the
final value of the baryon asymmetry.


(ii) {\it $\Delta L = 1$ scatterings with top quark.}  A comparison
between the corrected and uncorrected rates of the top-quark scattering $\gamma^{\rm top}_{H_{s}}\equiv \gamma
(q_3\,\bar t \leftrightarrow\ell\,N_1)$ with the Higgs exchanged in
the $s$-channel, and of the sum of the $t$- and $u$-channel scatterings
$\gamma^{\rm top}_{H_{t+u}}\equiv\gamma(q_3\, N_1 \leftrightarrow
\ell\, t) +\gamma(\bar t\, N_1 \leftrightarrow \ell\, \bar q_3)$ 
shows that the only corrections appearing at low temperatures, and
thus more relevant, are for $\gamma^{\rm top}_{H_{t+u}}$ (see
Figure 7.1 in Ref. \cite{Davidson:2008bu}).  They reduce the scattering
rates and suppress the related washouts.
This peculiar situation arises from the fact that in the zero
temperature limit there is a large enhancement $\sim
\ln(M_{N_1}/m_H)$ from the quasi-massless Higgs exchanged in the
$t$- and $u$-channels, which disappears when the Higgs
thermal mass $m_H(T)\sim T\sim M_{N_1}$ is included.

(iii) {\it $\Delta L = 1$ scatterings with the gauge bosons.} Here the
inclusion of thermal masses is required to avoid IR divergences that
would arise when massless $\ell$ (and $H$) states are exchanged in
the $t$- and $u$-channels. A naive use of some cutoff for the phase
space integrals to control the IR divergences can yield incorrect
estimates of the gauge bosons scattering rates~\cite{Giudice:2003jh})
and would be particularly problematic at low temperatures, where gauge
bosons scatterings dominate over top-quark scatterings.

%
%
%

\subsubsection{CP asymmetries}
\label{sec-08:thermalcp}
CP asymmetries arise from the interference of tree level and one-loop
amplitudes when the couplings involved have complex phases and the
loop diagrams have an absorptive part.  This last requirement is
satisfied whenever the loop diagram can be cut in such a way that the
particles in the cut lines can be produced on shell. For the CP
asymmetry in decay (at zero temperature) this is guaranteed by the
fact that the Higgs and the lepton final states coincide with the
states circulating in the loops.  However, in the hot plasma in which
$N_1$ decays occur, the Higgs and the lepton doublets are in thermal
equilibrium and their interactions with the background 
introduce in the CP asymmetries a dependence on the temperature 
$\epsilon\to\epsilon(T)$ that arises 
from various effects:
\begin{itemize} \itemsep -1pt

\item[\it i)] Absorption and re-emission of the
  loop particles by the medium require the use of finite temperature
  propagators.

\item[\it ii)] Stimulation of decays into bosons and blocking of
  decays into fermions in the dense background require proper
  modification of the final states density distributions. 

\item[\it iii)] Thermal motion of the decaying $N$'s with respect to
  the background breaks the Lorentz symmetry and affects the
  evaluation of the CP asymmetries.

\item[\it iv)] Thermal masses should be included in the finite
  temperature resummed propagators, and they also modify the fermion and
  boson dispersion relations.  Their inclusion yields the most
  significant modifications to the zero temperature results for the
  CP asymmetries.

\end{itemize}
The first three effects were investigated in Ref. \cite{Covi:1997dr} while
the effects of thermal masses was included in Ref. \cite{Giudice:2003jh}.
In principle, at finite temperature, there are additional effects
related to new cuts that involve the heavy $N_{2,3}$ neutrino lines.
These new cuts appear because the heavy particles in the loops may
absorb energy from the plasma and go on-shell.  However, for
hierarchical spectrum, $M_{2,3}\gg M_1$, the related effects are
Boltzmann suppressed by $\exp(-M_{2,3}/T)$ that at $T\sim M_1$ is a
tiny factor.  For a non-hierarchical spectrum, the effect of these new
cuts can however be sizable.  A detail study can be found in
Ref.~\cite{Garbrecht:2010sz}.

\subsubsection{Propagators and statistical distributions}
\label{sec-08:stat}
Particle propagators at finite temperature are computed in the real
time formalism of thermal field theory
\cite{Bellac:1996,Landsman:1986uw}.  In this formalism, ghost fields
dual to each of the physical fields have to be introduced, and
consequently the thermal propagators have $2\times2$ matrix
structures. For the one-loop computations of the absorptive parts of
the Feynman diagrams, the relevant propagator components are just
those of the physical lepton and Higgs fields. The
usual zero temperature propagators $-i S^0_\ell(p,m_\ell)=(p\!\!\!\slash
-m_\ell+i0^+)^{-1}$ and $-i D^0_H(p,m_H)=(p^2-m_H^2+i0^+)^{-1}$
acquire an additive  term that is proportional to the 
particle density distribution $n_{\ell,H}= [\exp(E_{\ell,H}/T)\pm 1]^{-1}$:
\begin{eqnarray}
\label{eq-08:thermalF}
\delta S_\ell(T) &=&
-2\pi\,n_\ell\,\left( p\!\!\!\slash -m_\ell \right)
\,\delta(p^2-m_\ell^2)\,,
\\
\label{eq-08:thermalB}
\delta D_H(T)&=&+2\pi\,  n_H
\,\delta(p^2-m_H^2)\,.
\end{eqnarray}
%
For the fermionic thermal propagators, there are other higher order
corrections (see Ref.~\cite{Giudice:2003jh}).  Unlike the case of
bosons, the interactions of the fermions with the thermal bath lead to
two different types of excitations with different dispersion
relations, that are generally referred to as `particles' and
`holes'~\cite{Giudice:2003jh}. The contributions of these two
fermionic modes were studied in
Refs.~\cite{Kiessig:2010pr,Kiessig:2011fw,Kiessig:2011ga} where it was
argued that in the strong washout regime they could give
non-negligible effects~\cite{Kiessig:2011ga}.
The leading effects in {\it i)} are proportional to the factor
$-n_\ell+n_H-2 n_\ell \, n_H$ that vanishes when the final states thermal
masses are neglected, because the Bose-Einstein and Fermi-Dirac
statistical distributions depend on the same argument,
$E_\ell=E_H=M_1/2$. As a consequence, the thermal corrections to the
fermion and boson propagators ($n_\ell$ and $n_H$) and the product of
the two thermal corrections ($n_\ell \, n_H$) cancel each other. This was interpreted 
as a complete compensation between stimulated emission
and Pauli blocking. As regards the effects in {\it ii)}, they lead to
overall factors that cancel between numerator and denominator in the
expression for the CP asymmetry\footnote{ 
A similar cancellation holds also in the supersymmetric case. However,
because of the presence of the superpartners $\tilde \ell\,,\tilde H$
both as final states and in the loops, the cancellation is more subtle
and it involves a compensation between the two types of corrections
{\it i)} and {\it ii)}.  We refer to Ref. \cite{Covi:1997dr} for details.}.
More recently, on the basis of a first principle derivation of the CP
asymmetry within a quantum BE approach (see Section
\ref{sec-08:quantum}) it has been claimed that the statistical factor
induced by thermal loops is instead $-n_\ell + n_H$, which does not
vanish even in the massless approximation. This would result in a
further enhancement in the CP asymmetry from the thermal
effects~\cite{Garny:2010nj}.

\subsubsection{Particle motion}
Given that the decaying particle $N_1$ is moving with respect to the
background (with velocity $\vec\beta$) the fermionic decay products
are preferentially emitted in the direction anti-parallel to the
plasma velocity (for which the thermal distribution is less occupied),
while the bosonic ones are emitted preferentially in the forward
direction (for which stimulated emission is more effective). Particle
motion then induces an angular dependence in the decay distribution at
order ${\cal O}(\beta)$.  In the total decay rate the ${\cal
  O}(\beta)$ anisotropy effect is integrated out, and only ${\cal
  O}(\beta^2)$ effects remain~\cite{Covi:1997dr}.  Therefore, while
accounting for thermal motion does modify the zero temperature
results, these corrections are numerically small
\cite{Covi:1997dr,Giudice:2003jh} and generally negligible.

\subsubsection{Thermal masses}
When the finite values of the light particle thermal masses are taken
into account, the arguments of the Bose-Einstein and Fermi-Dirac
statistical distributions are different. It is a good approximation
\cite{Giudice:2003jh} to use for the particle energies
$E_{\ell,H}=M_1/2\mp(m^2_H-m^2_\ell)/2M_1$. Since now
$E_\ell\neq E_H$, the prefactor $-n_\ell+n_H-2 n_\ell n_H$
that multiplies the thermal corrections does not vanish anymore, and
sizable corrections become possible.  The most relevant effect is
that the CP asymmetry vanishes when, as the temperature increases, the
sum of the light particles thermal masses approaches
$M_1$~\cite{Giudice:2003jh}. This is not surprising, since the
particles in the final state coincide with the particles in the loop,
and therefore when the decay becomes kinematically forbidden, also the
particles in the loop cannot go on the mass shell. 
When the temperature is large enough that $m_H(T)> m_\ell(T)+M_1$ the
Higgs can decay, and then there is a new source of lepton number
asymmetry associated with $H \to \ell N_1$.  The CP asymmetry in Higgs
decays $\epsilon_H$ can be up to one order of magnitude larger than
the CP asymmetry in $N_1$ decays~\cite{Giudice:2003jh}.  While
this could represent a dramatic enhancement of the CP asymmetry,
$\epsilon_H$ is non-vanishing only at temperatures $T \gsim T_H \sim 5
M_1$, when the kinematic condition for its decays is
satisfied.  Therefore, in the strong washout regime, no trace of this
effect survives. On the other hand, rather large couplings
are required in order that Higgs decays can occur before the phase
space closes: the decay rate can attain thermal equilibrium only when
$\tilde m \gsim (T_H/M_1)^2 m_*\gg m_*$, and therefore, in the weak
washout regime ($\tilde m \lsim m_*$), these decays always remain
strongly out of equilibrium.  This means that only a small fraction of
the Higgs particles have actually time to decay, and the
lepton-asymmetry generated in this way is accordingly suppressed.

In summary, while the corrections to the CP asymmetries can be
significant at $T\gsim M_1$ (and quite large at $T\gg M_1$ for Higgs
decays), in the low temperature regime, where the precise value of
$\epsilon$ plays a fundamental role in determining the final value of
the baryon asymmetry, there are almost no effects, and the zero
temperature results still give a reliable approximation.

\subsection{Decays of the heavier right-handed neutrinos}
\label{sec-08:heavier}

In leptogenesis studies, the effects of $N_{2,3}$ are often neglected,
which in many cases is not a good approximation.  This is obvious for
example when $N_1$ dynamics is irrelevant for leptogenesis:
$\epsilon_1 \ll 10^{-6}$ cannot provide enough CP asymmetry to account
for baryogenesis, and $\tilde m_1 \ll m_*$ implies that $N_1$ washout
effects are negligible. It is then clear that any asymmetry generated
in $N_{2,3}$ decays can survive, and becomes crucial for the success
of leptogenesis.  Another case in which it is intuitively clear that
$N_{2,3}$ effects can be important, is when the RH neutrino spectrum
is compact, which means that $M_{2,3}$ have values within a factor of
a few from $M_1$. Then $N_1$ and $N_{2,3}$ contributions to
leptogenesis can be equally important and must be summed up.  A model
with compact RH neutrino spectrum in which $N_{2,3}$ dynamics is of
crucial importance was recently discussed in Ref. \cite{Buccella:2012kc}.

It is less obvious that $N_{2,3}$ effects can also be important for 
a hierarchical RH spectrum and when $N_1$ is strongly coupled.  This
can happen because decoherence effects related to $N_1$-interactions
can project the asymmetry generated in $N_{2,3}$ decays onto a
flavour direction that remains protected against $N_1$
washouts~\cite{Barbieri:1999ma,Nielsen:2001fy,Strumia:2006qk,Engelhard:2006yg}.
Let us illustrate this with an example. Let us assume that 
a sizable asymmetry is generated in $N_2$ decays, while $N_1$
leptogenesis is inefficient and fails, that is:
\begin{equation}\label{eq-08:contms}
\tilde m_2\not\gg m_*,\ \ \ \ \tilde m_1\gg m_*.
\end{equation}
Assuming  also a strong hierarchy and that leptogenesis occurs  
 thermally guarantees that~\cite{Engelhard:2006yg}: 

\begin{equation}
n_{N_1}(T\sim M_2)\approx 0,\ \ \ n_{N_2}(T\sim M_1)\approx 0.
\end{equation}
Thus, the dynamics of $N_2$ and $N_1$ are decoupled: there are neither
$N_1$-related washout effects during $N_2$ leptogenesis, nor
$N_2$-related washout effects during $N_1$ leptogenesis.
The $N_2$ decays into a combination of lepton doublets that we denote
by $\ell_2$:
\begin{equation}
|\ell_2\rangle=(Y^\dagger Y)_{22}^{-1/2}
\sum_\alpha Y_{\alpha 2}|\ell_\alpha\rangle.
\end{equation}
The second condition in eq. (\ref{eq-08:contms}) implies that already at
$T\gsim M_1$ the $N_1$-Yukawa interactions are sufficiently fast to
quickly destroy the coherence of $\ell_2$. Then a statistical mixture
of $\ell_1$ and of the state orthogonal to $\ell_1$ builds up, and it
can be described by a suitable diagonal density matrix. 
Let us consider the simple case where both $N_2$ and $N_1$ decay at
$T\gsim10^{12}\>$ GeV, so that flavour effects are irrelevant.
A convenient choice for an orthogonal basis for the lepton doublets is
$(\ell_1,\ell_0,\ell_0^\prime)$ where, without loss of generality,
$\ell_0^\prime$ satisfies $\langle\ell_0^\prime|\ell_2\rangle=0$. Then
the asymmetry $\Delta Y_{\ell_2}$ produced in $N_2$ decays decomposes into
two components:
\begin{equation}
\Delta Y_{\ell_0}=c^2\,\Delta Y_{\ell_2},\qquad \ \Delta
Y_{\ell_1}=s^2\,\Delta Y_{\ell_2},
\end{equation}
where $c^2\equiv|\langle\ell_0|\ell_2\rangle|^2$ and $s^2=1-c^2$. The
crucial point here is that we expect, in general, $c^2\neq0$ and,
since $\langle\ell_0|\ell_1\rangle=0$, $\Delta Y_{\ell_0}$ is protected
against $N_1$ washout. Consequently, a finite part of the asymmetry
$\Delta Y_{\ell_2}$ from $N_2$ decays survives through $N_1$ leptogenesis. A
more detailed analysis \cite{Engelhard:2006yg} finds that $\Delta Y_{\ell_1}$
is not entirely washed out, resulting in the final lepton asymmetry 
$Y_{\Delta L}=(3/2)\Delta Y_{\ell_0}=(3/2)c^2\,\Delta Y_{\ell_2}$.

For $10^9\,$GeV$\lsim M_1\lsim 10^{12}\,$GeV, flavour issues modify
the quantitative details, but the qualitative picture, and in
particular the survival of a finite part of $\Delta Y_{\ell_2}$, still
holds. In contrast, for $M_1\lsim 10^9\,$GeV, the full flavour basis
$(\ell_e,\ell_\mu,\ell_\tau)$ is resolved and thus there are
no directions in flavour space where an asymmetry is protected, so that 
$Y_{\ell_2}$ can be erased entirely. A dedicated study in which the
various flavour regimes for $N_{1,2,3}$ decays are considered can be
found in Ref. \cite{Blanchet:2011xq}.

In conclusion $N_{2,3}$ leptogenesis cannot be ignored,
unless one of the following conditions holds:
\begin{enumerate} \itemsep -2pt
\item The reheat temperature is below $M_2$.
\item The asymmetries and/or the washout factors vanish,
  $\epsilon_{N_2}\eta_2\approx0$ and $\epsilon_{N_3}\eta_3\approx0$.
\item $N_1$-related washout is still significant at $T\lsim10^9\,$GeV.
\end{enumerate}
%


\subsection{Quantum Boltzmann equations}
\label{sec-08:quantum}

So far we have analyzed the leptogenesis dynamics by adopting the
classical BE of motion. An interesting question which
has attracted some attention recently \cite{Buchmuller:2000nd,
DeSimone:2007rw,Garny:2009rv,Garny:2009qn,Anisimov:2010aq,Garny:2010nj,
Beneke:2010wd,Beneke:2010dz,Garny:2010nz,Garbrecht:2010sz,Anisimov:2010dk}
is under which circumstances the classical BE can be
safely applied to get reliable results and, conversely, when a more
rigorous quantum approach is needed.  Quantum BE are
obtained starting from the non-equilibrium quantum field theory based
on the Closed Time-Path (CTP) formulation \cite{chou:1984es}.  Both,
CP violation from wave-function and vertex corrections are
incorporated. Unitarity issues are resolved and an accurate account
of all quantum-statistical effects on the asymmetry is made. Moreover,
the formulation in terms of Green functions bears the potential of
incorporating corrections from Thermal Field Theory within the CTP
formalism.

In the CTP formalism, particle number densities are replaced by
Green's functions obeying a set of equations which, under some
assumptions, can be reduced to a set of kinetic equations describing
the evolution of the lepton asymmetry and the RH neutrinos. These
kinetic equations are non-Markovian and present memory effects. In
other words, differently from the classical approach where every
scattering in the plasma is independent from the previous one, the
particle abundances at a given time depend upon the history of the
system. The more familiar energy-conserving delta functions are
replaced by retarded time integrals of time-dependent kernels and
cosine functions whose arguments are the energy involved in the
various processes. Therefore, the non-Markovian kinetic equations
include the contribution of coherent processes throughout the history
of the kernels and the relaxation times are expected to be typically
longer than the one dictated by the classical approach.

If the time range of the kernels are shorter than the
relaxation time of the particles abundances, the solutions to the
quantum and the classical BE differ only by terms of the order of the ratio
of the timescale of the kernel to the relaxation timescale of the
distribution. In thermal leptogenesis this is typically the case. However, 
there are situations where this does not happen. For instance, in the
case of resonant leptogenesis, 
two RH (s)neutrinos $N_1$ and $N_2$ 
 are almost degenerate
in mass and the CP asymmetry from the decay of the first RH neutrino $N_1$ 
is resonantly enhanced if the mass difference
$\Delta M=(M_2-M_1)$
is of the order of the decay rate of the second RH neutrino
$\Gamma_{N_2}$ . The typical timescale
to build up coherently the CP asymmetry is of the order of $1/\Delta M$, which
can be larger than the timescale $\sim 1/\Gamma_{N_1}$ 
for the change of the abundance of the
$N_1$'s.

Since  we
need the time evolution of the particle asymmetries with definite 
initial conditions and not
simply the transition amplitude of particle reactions, 
the ordinary equilibrium quantum field theory at finite temperature   
is not the appropriate tool. 
The most appropriate extension of the field theory
to deal with non-equilibrium phenomena amounts to generalizing
 the time contour of
integration to a closed time-path. More precisely, the time integration
contour is deformed to run from $t_0$ to $+\infty$ and back to
$t_0$. 
The CTP  formalism  is a powerful 
Green's function
formulation for describing non-equilibrium phenomena in field theory.  It
allows to describe phase-transition phenomena and to obtain a
self-consistent set of quantum BE.
The formalism yields various quantum averages of
operators evaluated in the in-state without specifying the out-state. 
On the contrary, the ordinary quantum field theory 
yields quantum averages of the operators evaluated  
with an in-state at one end and an out-state at the other. 

For example, because of the time-contour deformation, the partition function 
in the in-in formalism for a complex scalar field is defined to be
\begin{eqnarray}
Z\left[ J, J^{\dagger}\right] &=& {\rm Tr}\:
\left[ \mathcal{T}\left( {\rm exp}\left[i\:\int_C\:\left(J(x)\phi(x)+
J^{\dagger}(x)\phi^{\dagger}(x) \right) d^4 x\right]\right)\rho\right]\nonumber\\
&=& {\rm Tr}\:\left[ \mathcal{T}_{+}\left( {\rm exp}\left[ i\:\int\:
\left(J_{+}(x)\phi_{+}(x)+J^{\dagger}_{+}(x)\phi^{\dagger}_{+}(x) \right) d^4 x\right]\right)
\right.
\nonumber\\
&\times&\left.  \mathcal{T}_{-}\left( {\rm exp}\left[
      -i\:\int\:\left(J_{-}(x)\phi_{-}(x)+J^{\dagger}_{-}(x)\phi^{\dagger}_{-}(x)
      \right) d^4 x\right]\right) \rho\right],
\end{eqnarray}
where $C$ in the integral denotes that the time integration contour
runs from $t_0$ to plus infinity and then back to $t_0$ again. The
symbol $\rho$ represents the initial density matrix and the fields are
in the Heisenberg picture and defined on this closed time-contour
(plus and minus subscripts refer to the positive and negative
directional branches of the time path, respectively). The
time-ordering operator along the path is the standard one ($\mathcal
T_+$) on the positive branch, and the anti-time-ordering ($\mathcal
T_-$) on the negative branch.  As with the Euclidean-time formulation,
scalar (fermionic) fields $\phi$ are still periodic (anti-periodic) in
time, but with $\phi(t,\vec{x})=\phi(t-i\beta,\vec{x})$, $\beta=1/T$.
The temperature $T$ appears due to boundary condition, and time is now
explicitly present in the integration contour.

We must now identify field
variables with arguments on the positive or negative directional
branches of the time path. This doubling of field variables leads to
six  different real-time propagators on the contour.  These six
propagators are not independent, but using all of them simplifies the 
notation. 
For a generic charged  scalar field $\phi$ they are defined as 
\begin{eqnarray}
\label{eq-08:def1}
G_{\phi}^{>}\left(x, y\right)&=&
-G_{\phi}^{-+}\left(x, y\right)=
-i\langle
\phi(x)\phi^\dagger (y)\rangle,\nonumber\\
G_{\phi}^{<}\left(x,y\right)&=&-G_{\phi}^{+-}\left(x, y\right)=-i\langle
\phi^\dagger (y)\phi(x)\rangle,\nonumber\\
G^t _{\phi}(x,y)&=& G_{\phi}^{++}\left(x, y\right)=
\theta(x,y) G_{\phi}^{>}(x,y)+\theta(y,x) 
G_{\phi}^{<}(x,y),\nonumber\\
G^{\overline{t}}_{\phi} (x,y)&=& G_{\phi}^{--}\left(x, y\right)=
\theta(y,x) G_{\phi}^{>}(x,y)+
\theta(x,y) G_{\phi}^{<}(x,y), \nonumber\\
G_{\phi}^r(x,y)&=&G_{\phi}^t-G_{\phi}^{<}=G_{\phi}^{>}-
G^{\overline{t}}_{\phi}, 
\:\:\:\: G_{\phi}^a(x,y)=G^t_{\phi}-G^{>}_{\phi}=G_{\phi}^{<}-
G^{\overline{t}}_{\phi},
\end{eqnarray}
where the last two Green's functions are the retarded and advanced 
Green's functions respectively and $\theta(x,y)\equiv\theta(t_x-t_y)$ 
is the step function.  

For a generic fermion field $\psi$ the six 
different propagators are analogously defined as
\begin{eqnarray}
\label{eq-08:def2}
G^{>}_{\psi}\left(x, y\right)&=&-G^{-+}_{\psi}\left(x, y\right)=-i\langle
\psi(x)\overline{\psi} (y)\rangle,\nonumber\\
G^{<}_{\psi}\left(x,y\right)&=&-G^{+-}_{\psi}\left(x, y\right)=+i\langle
\overline{\psi}(y)\psi(x)\rangle,\nonumber\\
G^{t}_{\psi} (x,y)&=& G^{++}_{\psi}\left(x, y\right)
=\theta(x,y) G^{>}_{\psi}(x,y)+
\theta(y,x) G^{<}_{\psi}(x,y),\nonumber\\
G^{\overline{t}}_{\psi} (x,y)&=& G^{--}_{\psi}\left(x, y\right)=
\theta(y,x) G^{>}_{\psi}(x,y)+
\theta(x,y) G^{<}_{\psi}(x,y),\nonumber\\
G^r_{\psi}(x,y)&=&G^{t}_{\psi}-G^{<}_{\psi}=G^{>}_{\psi}
-G^{\overline{t}}_{\psi}, \:\:\:\: G^a_{\psi}(x,y)=G^{t}_{\psi}-
G^{>}_{\psi}=G^{<}_{\psi}-G^{\overline{t}}_{\psi}.
\end{eqnarray}
From the definitions of the Green's functions, one can see
that  the hermiticity properties

\begin{equation}
\label{eq-08:prop}
\left(i\gamma^0 G_\psi(x,y)\right)^\dagger=i\gamma^0  G_\psi(y,x), \,\,\,
\left(i G_\phi(x,y)\right)^\dagger=i G_\phi(y,x), 
\end{equation}
are satisfied.
For interacting systems,
whether in equilibrium or not, one must 
define and calculate self-energy functions. 
Again, there are six of them: $\Sigma^{t}$, 
$\Sigma^{\overline{t}}$, $\Sigma^{<}$, $\Sigma^{>}$, 
$\Sigma^r$ and $\Sigma^a$. The same 
relationships exist among them as for the 
Green's functions in  (\ref{eq-08:def1}) and (\ref{eq-08:def2}), such as
\begin{equation}
\Sigma^r=\Sigma^{t}-\Sigma^{<}=\Sigma^{>}-\Sigma^{\overline{t}}, 
\:\:\:\:\Sigma^a=\Sigma^{t}-\Sigma^{>}=\Sigma^{<}-\Sigma^{\overline{t}}. 
\end{equation}
The self-energies are incorporated into the Green's 
functions through the use of  Dyson's equations. 
A useful notation may be introduced which expresses 
four of the six Green's functions as the elements of 
two-by-two matrices 

\begin{equation}
\widetilde{G}=\left(
\begin{array}{cc}
G^{t} & \pm G^{<}\\
G^{>} & - G^{\overline{t}}
\end{array}\right), \:\:\:\:
\widetilde{\Sigma}=\left(
\begin{array}{cc}
\Sigma^{t} & \pm \Sigma^{<}\\
\Sigma^{>} & - \Sigma^{\overline{t}}
\end{array}\right),
\end{equation}
where the upper signs refer to the bosonic case and the lower signs 
to the fermionic case. For systems either in equilibrium or in non-equilibrium, 
Dyson's equation is most easily expressed by using the matrix notation
\begin{equation}
\label{eq-08:d1}
\widetilde{G}(x,y)=\widetilde{G}^0(x,y)+\int d^4 z_1
\int d^4 z_2 \: \widetilde{G}^0(x,z_1)
\widetilde{\Sigma}(z_1,z_2)\widetilde{G}(z_2,y),
\end{equation}
where the superscript ``0'' on the Green's functions means 
to use those for noninteracting system.   It is useful 
to notice that Dyson's equation can be written in an 
alternative form, instead of  (\ref{eq-08:d1}), with $\widetilde{G}^0$ 
on the right in the interaction terms,
\begin{equation}
\label{eq-08:d2}
\widetilde{G}(x,y)=\widetilde{G}^0(x,y)+\int\: d^4z_3\:\int d^4z_4\: 
\widetilde{G}(x,z_3)
\widetilde{\Sigma}(z_3,z_4)\widetilde{G}^0(z_4,y).
\end{equation}
Eqs.~(\ref{eq-08:d1}) and (\ref{eq-08:d2}) are the 
starting points to derive the quantum BE
describing the time evolution of the 
CP-violating particle density asymmetries.

To proceed, one has to choose a form for  the propagators. 
For a generic fermion $\psi$ (and similarly for scalars) 
one may adopt the real-time propagator in the form  
$G^{t}_\psi({\bf k},t_x-t_y)$ in terms of the spectral
function $\rho_\psi({\bf k},k_0)$ 
\bea
G^{t}_\psi({\bf k},t_x-t_y)&=&\int_{-\infty}^{+\infty}\:
\frac{d k^0}{2\pi}\:e^{-i k^0(t_x-t_y)}\:\rho_\psi({\bf k},k^0)\nonumber\\
&\times&\left\{
\left[1-f_\psi(k^0)\right]\theta(t_x-t_y)-f_\psi(k^0)\theta(t_y-t_x)\right\},
\label{eq-08:rho1}
\eea
where $f_\psi(k^0)$ represents the fermion distribution function.
Again, particles must be
substituted by quasiparticles,  dressed propagators are to be adopted
and 
 self-energy
corrections to
the propagator modify the dispersion relations by 
introducing a finite width $\Gamma_\psi(k)$. 
 For a fermion with 
chiral mass $m_\psi$, one may safely choose
\begin{equation}
\rho_\psi({\bf k},k^0) =  i\left(\not  k +m_\psi\right)
\left[\frac{1}{(k^0+i\epsilon+ i
\Gamma_{\psi})^2-\omega_{\psi}^2(k)}-
\frac{1}{(k^0-i\epsilon-i\Gamma_{\psi})^2
-\omega_{\psi}^2(k)}\right],
\label{eq-08:rofermion}
\end{equation}
where  $\omega_{\psi}^2(k)={\bf k}^2 +
M_{\psi}^2(T)$ and $M_{\psi}(T)$
is the effective thermal mass of the fermion in the plasma (not a
chiral mass). 
Performing the integration over $k^0$ and picking up 
the poles of the spectral function (which is valid for 
quasi-particles in equilibrium or very close to equilibrium), 
one gets 
\begin{eqnarray}
G_\psi^{>}({\bf k},t_x-t_y)&=&-\frac{i}{2\omega_\psi}
\left\{
\left(\not k+ m_\psi\right) 
\left[1-f_\psi(\omega_\psi-i\Gamma_\psi)\right]\:
e^{-i(\omega_\psi-i\Gamma_\psi)(t_x-t_y)}\right.\nonumber\\
&+&\left.\gamma^0\left(\not k- 
m_\psi\right) \gamma^0
\overline{f}_\psi(\omega_\psi+i\Gamma_\psi)\:
e^{i(\omega_\psi+i\Gamma_\psi)(t_x-t_y)}\right\},\nonumber\\
G_\psi^{<}({\bf k},t_y-t_x)&=&\frac{i}{2\omega_\psi}\left\{
\left(\not k+ m_\psi\right)
f_\psi(\omega_\psi+i\Gamma_\psi)\:
e^{-i(\omega_\psi-i\Gamma_\psi)(t_x-t_y)}\right.\nonumber\\
&+&\left.\gamma^0\left(\not k- m_\psi\right)
\gamma^0
\left[1-\overline{f}_\psi(\omega_\psi-i\Gamma_\psi)\right]\:
e^{i(\omega_\psi+i\Gamma_\psi)(t_x-t_y)}\right\},  
\label{eq-08:b}
\end{eqnarray}
where $k^0=\omega_\psi$ and  
$f_\psi, \overline{f}_\psi$ denote the distribution function of
the fermion particles and antiparticles, respectively. 
The expressions (\ref{eq-08:b}) are valid for $t_x-t_y>0$.

The above definitions
hold for the  lepton doublets (after inserting the
chiral LH projector $P_L$), as well as for the Majorana
RH neutrinos, for which one has to assume identical  particle and antiparticle
distribution functions and insert the inverse of the charge conjugation matrix 
$C$ in the dispersion relation. 

To elucidate  further the impact of the CTP approach and to see under which 
conditions one can obtain the classical BE from the quantum ones,  
one may consider  the dynamics of the lightest RH neutrino $N_1$. 
To find its quantum BE we start from eq.~(\ref{eq-08:d1}) for the Green's function $G^<_{N_1}$ 
of the RH neutrino
$N_1$

\begin{eqnarray}
\label{eq-08:vv}
\left(i\stackrel{\rightarrow}{\not  \partial}_x -M_1\right)
G^{<}_{N_1}(x,y)&=&- 
\int d^4 z \,\left[-\Sigma^{t}_{N_1}(x,z)
G^<_{N_1}(z,y)+\Sigma^{<}_{N_1}(x,z)
G^{\overline{t}}_{N_1}(z,y)
\right]\nonumber\\
&=&\int\: d^3 z\:\int_{0}^{t}\: dt_z\:
\left[\Sigma^{>}_{N_1}(x,z)
G^<_{N_1}(z,y)-\Sigma^{<}_{N_1}(x,z)
G^{>}_{N_1}(z,y)
\right].\nonumber\\
&&
\end{eqnarray}
Adopting the corresponding form for the RH neutrino propagator and the center-of-mass coordinates
\begin{equation}
\label{eq-08:dd}
X\equiv(t,\vec{X})\equiv\frac{1}{2}(x+y),\:\:\:\: (\tau,\vec{r})\equiv x-y,
\end{equation}
one ends up with the following equation 

\begin{eqnarray}
\frac{\partial f_{N_1}({\bf k},t) }{\partial t}&=&
-2
\int_{0}^{t} dt_z\, \int\frac{d^3{\bf p}}{(2\pi)^3}\, 
\frac{1}{2\omega_{\ell}({\bf p})}\frac{1}{2\omega_H({\bf k}-{\bf p})}
\frac{1}{\omega_{N_1}({\bf k})}
\left|{\cal M}(N_1\rightarrow {\ell} H)\right|^2
\nonumber\\
&& \times \left[
f_{N_1}({\bf k},t)(1-f_{{\ell}}({\bf p},t))(1+f_{H}({\bf k}-{\bf p},t))
\right.\nonumber\\
&&\left.
-f_{{\ell}}({\bf p},t)f_{H}({\bf k}-{\bf p},t)(1-f_{N_1}({\bf k},t))
\right]\nonumber\\
 && \times \cos\left[\left(\omega_{N_1}({\bf k})-
\omega_{\ell}({\bf p})-
\omega_H({\bf k}-{\bf p})\right)(t-t_z)\right]\nonumber\\
&\simeq& -2
\int_{0}^{t} dt_z\, \int\frac{d^3{\bf p}}{(2\pi)^3}\, 
\frac{1}{2\omega_{\ell}({\bf p})}\frac{1}{2\omega_H({\bf k}-{\bf p})}
\frac{1}{\omega_{N_1}({\bf k})}\left|{\cal M}(N_1\rightarrow {\ell} H)
\right|^2\nonumber\\
&& \times\left(
f_{N_1}({\bf k},t)-f^{eq}_{{\ell}}({\bf p})f^{eq}_{H}({\bf k}-
{\bf p})
\right)\nonumber\\
 && \times\cos\left[\left(\omega_{N_1}({\bf k})-
\omega_{\ell}({\bf p})-
\omega_H({\bf k}-{\bf p})\right)(t-t_z)\right].
\label{eq-08:cy}
\end{eqnarray}
This equations holds under the assumption that the relaxation
timescale for the distribution functions are longer than the timescale
of the non-local kernels so that they can be extracted out of the time
integral. This allows to think the distributions as functions of the
center-of-mass time only.  We have set to zero the damping rates of
the particles in eq.~(\ref{eq-08:b}) and retained only those cosines giving
rise to energy delta functions that can be satisfied.  Under these
assumptions, the distribution function may be taken out of the time
integral, leading -- at large times -- to the so-called Markovian
description. The kinetic equation (\ref{eq-08:cy}) has an obvious
interpretation in terms of gain minus loss processes, but the retarded
time integral and the cosine function replace the familiar energy-conserving 
delta functions. In the second passage, we have also made
the usual assumption that all distribution functions are smaller than
unity and that those of the Higgs and lepton doublets are in
equilibrium and much smaller than unity, $f_{\ell} f_H\simeq
f^{eq}_{{\ell}}f^{eq}_{H}$. Elastic scatterings are typically fast
enough to keep kinetic equilibrium.  For any distribution function we
may write $f=(n/n^{eq})f^{eq}$, where $n$ denotes the total number
density.  Therefore, eq.~(\ref{eq-08:cy}) can be re-written as

\begin{eqnarray}
\frac{\partial n_{N_1}}{\partial t}&=&
-\langle \Gamma_{N_1}(t)\rangle n_{N_1}+\langle 
\widetilde{\Gamma}_{N_1}(t)\rangle n^{eq}_{N_1},\nonumber\\
\langle\Gamma_{N_1}(t)\rangle&=&\int_{0}^{t} dt_z
\int\frac{d^3{\bf k}}{(2\pi)^3}
\frac{f^{eq}_{N_1}}{n^{eq}_{N_1}}
\,\Gamma_{N_1}(t),\nonumber\\
\Gamma_{N_1}(t)&=&
2
\int\frac{d^3{\bf p}}{(2\pi)^3}\, 
\frac{\left|{\cal M}(N_1\rightarrow {\ell} 
H)\right|^2}{2\omega_{\ell} 2\omega_H\omega_{N_1}}
\cos\left[\left(\omega_{N_1}-
\omega_{\ell}-
\omega_H\right)(t-t_z)\right],  \nonumber\\
\langle\widetilde{\Gamma}_{N_1}(t)\rangle&=&\int_{0}^{t} dt_z
\int\frac{d^3{\bf k}}{(2\pi)^3}
\frac{f^{eq}_{N_1}}{n^{eq}_{N_1}}
\,\widetilde{\Gamma}_{N_1}(t),\nonumber\\
\widetilde{\Gamma}_{N_1}(t)&=&
2
\int\frac{d^3{\bf p}}{(2\pi)^3}\, 
\frac{f^{eq}_{\ell}f^{eq}_H}{f^{eq}_{N_1}}
\frac{\left|{\cal M}(N_1\rightarrow {\ell} 
H)\right|^2}{2\omega_{\ell} 2\omega_H\omega_{N_1}}
\cos\left[\left(\omega_{N_1}-
\omega_{\ell}-
\omega_H\right)(t-t_z)\right],  \nonumber\\
\label{eq-08:ddd}
\end{eqnarray}
where $\langle \Gamma_{N_1}(t)\rangle$ is the time-dependent 
thermal average of the
Lorentz-dilated decay width. 
Integrating over
large times, $t\rightarrow \infty$, thereby replacing the
cosines by energy-conserving delta functions\

\begin{equation}
\int_{0}^{\infty}dt_z\,\cos\left[\left(\omega_{N_1}-
\omega_{\ell}-
\omega_H\right)(t-t_z)\right]=\pi\delta\left(
\omega_{N_1}-
\omega_{\ell}-
\omega_H
\right),
\end{equation}
we find that the two averaged rates $\langle\Gamma_{N_1}\rangle$ and
$\langle\widetilde{\Gamma}_{N_1}\rangle$ coincide and we recover 
the usual classical BE for the RH distribution
function

\begin{eqnarray}
\frac{\partial n_{N_1}}{\partial t}&=&-\langle\Gamma_{N_1}\rangle\left(
n_{N_1}-n^{eq}_{N_1}
\right),\nonumber\\
\langle\Gamma_{N_1}\rangle&=&
\int\frac{d^3{\bf k}}{(2\pi)^3}
\frac{f^{eq}_{N_1}}{n^{eq}_{N_1}}
\int\frac{d^3{\bf p}}{(2\pi)^3}\, 
\frac{\left|{\cal M}(N_1\rightarrow {\ell} H)
\right|^2}{2\omega_{\ell}2\omega_H\omega_{N_1}}\,(2\pi)\delta\left(
\omega_{N_1}-
\omega_{\ell}-
\omega_H\right).\nonumber\\
&&
\end{eqnarray}
Taking the time interval to infinity, namely implementing Fermi's golden rule,
results in neglecting memory effects, which in turn results only
in on-shell processes contributing to the rate equation. The main difference
between the classical and the quantum BE can be traced to
memory effects and to the fact that the time evolution
of the distribution function is
non-Markovian. The memory of the past time evolution 
translates into off-shell processes. 

Similarly, 
one can show that the equation obeyed by the asymmetry reads
\begin{eqnarray}
\label{eq-08:aa}
 \frac{\partial n_{{\Delta L}_\alpha}(X)}{\partial t}
&=&-\int\: d^3 z\:\int_{0}^{t} dt_z\:{\rm Tr}
\left[\Sigma^{>}_{{\ell_\alpha}}(X,z) G^{<}_{{\ell_\alpha}}(z,X)
-  G^{>}_{{\ell_\alpha}}(X,z) \Sigma^{<}_{{\ell_\alpha}}(z,X) \right.\nonumber\\  
&+&\left.  G^{<}_{{\ell_\alpha}}(X,z)\Sigma^{>}_{{\ell_\alpha}}(z,X)-\Sigma^{<}_{{\ell_\alpha}}(X,z) 
G^{>}_{{\ell_\alpha}}(z,X)\right].
\end{eqnarray}
Proceeding as for the RH neutrino equation one finds (including for
the moment only the 1-loop wave contribution to the CP asymmetry
$\epsilon^{\alpha}_{w}$)
\begin{eqnarray}
\frac{\partial n_{{\Delta L}_\alpha}}{\partial t}
&=&
\epsilon^{\alpha}_{w}(t)
\langle\Gamma_{N_1}\rangle\left(
n_{N_1}-n^{eq}_{N_1}
\right),\nonumber\\
\epsilon^{\alpha}_{w}(t)
&=&
-\frac{4}{\langle\Gamma_{N_1}\rangle}\sum_{\beta=1}^3{\rm Im}
\left(Y_{1\alpha}
Y_{1\beta}Y^\dagger_{\beta 2}
Y^\dagger_{\alpha 2}\right)\nonumber\\
&\times&\int_0^t dt_z\int_0^{t_z} dt_2\int_0^{t_2} dt_1
e^{-\Gamma_{N_2}(t_z-t_2)}
e^{-\left(\Gamma_{\ell_\beta}+\Gamma_H\right)(t_2-t_1)}
\int\frac{d^3{\bf k}}{(2\pi)^3}
\frac{f^{eq}_{N_1}}{n^{eq}_{N_1}}\nonumber\\
&\times&
\int\frac{d^3{\bf p}}{(2\pi)^3}
\frac{1-f_{\ell_\beta}^{eq}({\bf p})+f_{H}^{eq}({\bf k}-{\bf p})
}{2\omega_{\ell_\beta}({\bf p})2\omega_H({\bf k}-{\bf p})
\omega_{N_1}({\bf k})}
\int\frac{d^3{\bf q}}{(2\pi)^3}
\frac{1-f_{\ell_\alpha}^{eq}({\bf q})+f_{H}^{eq}({\bf k}-{\bf q})
}{2\overline{\omega}_{\ell_\alpha}({\bf q})2\overline{\omega}_H({\bf k}
-{\bf q})
\omega_{N_2}({\bf k})}\nonumber\\
&\times&
 {\rm sin}\left(\omega_{N_1}(t-t_1)+\left(
\omega_{\ell_\beta}+\omega_H\right)(t_1-t_2)+\omega_{N_2}(t_2-t_z)
+\left(
\overline{\omega}_{\ell_\alpha}+\overline{\omega}_H\right)(t_z-t)
\right)\nonumber\\
&\times&
{\rm Tr}\left(M_1 P_L \not p  M_2 \not q\right), 
\end{eqnarray}
where, to avoid double counting, 
we have not inserted the decay rates in the propagators of the initial
and final states and, 
for simplicity, we have assumed that the damping rates of the 
lepton doublets and the Higgs field are constant in time. This should
be a good approximation as the damping rate are to be computed 
for momenta of order of the mass of the RH neutrinos. 
As expected from first principles, we find that the
CP asymmetry is a function of time and its value at a given
instant depends upon the previous history of the system. 

Performing the time integrals and retaining only those pieces which
eventually give rise to energy-conserving delta functions in the 
Markovian limit,  we obtain
\begin{eqnarray}
\label{eq-08:eeee}
\epsilon^{\alpha}_{w}(t)&=&
-\frac{4}{\langle\Gamma_{N_1}\rangle}\sum_{\beta =1}^3{\rm Im}
\left(Y_{1\alpha}
Y_{1\beta}Y^\dagger_{\beta 2}
Y^\dagger_{\alpha 2}\right)\nonumber\\
&\times&
\int_0^t dt_z
\frac{\cos\left[\left(\omega_{N_1}-
\overline{\omega}_{\ell_\alpha}-
\overline{\omega}_H\right)(t-t_z)\right]}{
\left(\Gamma_{N_2}^2+(\omega_{N_2}-\omega_{N_1})^2\right)
\left((\Gamma_{\ell_\beta}+\Gamma_{H})^2+(\omega_{N_1}-
\omega_{\ell_\beta}-\omega_H
)^2\right)}
\nonumber\\
&\times&
\int\frac{d^3{\bf k}}{(2\pi)^3}
\frac{f^{eq}_{N_1}}{n^{eq}_{N_1}}
(\Gamma_{\ell_\beta}+\Gamma_{H})\Bigg(2\,(\omega_{N_2}-\omega_{N_1})
\,{\rm sin}^2 \left[\frac{(\omega_{N_2}-\omega_{N_1})t_z}{2}\right]
\nonumber\\
&-&
\Gamma_{N_2}\,{\rm sin} \left[(\omega_{N_2}-\omega_{N_1})t_z\right]
\Bigg)
\int\frac{d^3{\bf p}}{(2\pi)^3}
\frac{1-f_{\ell_\beta}^{eq}({\bf p})+f_{H}^{eq}({\bf k}-{\bf p})
}{2\omega_{\ell_\beta}({\bf p})2\omega_H({\bf k}-{\bf p})
\omega_{N_1}({\bf k})}
\nonumber\\
&\times&
\int\frac{d^3{\bf q}}{(2\pi)^3}
\frac{1-f_{\ell_\alpha}^{eq}({\bf q})+f_{H}^{eq}({\bf k}-{\bf q})
}{2\overline{\omega}_{\ell_\alpha}({\bf q})2\overline{\omega}_H({\bf k}
-{\bf q})
\omega_{N_2}({\bf k})}\;
{\rm Tr}\left(M_1 P_L \not p M_2 \not q\right).
\end{eqnarray}
From this expression it is already 
manifest that the typical timescale for the
building up of the coherent CP asymmetry depends crucially on
the difference in energy of the two RH neutrinos.

If we now let the upper limit of the time integral to take large
values, we neglect the memory effects, the CP asymmetry picks
contribution only from the on-shell processes. Taking the damping
rates of the lepton doublets equal for all the flavours and the RH
neutrinos nearly at rest with respect to the thermal bath, the CP
asymmetry reads (now summing over all flavour indices)
\begin{eqnarray}
\epsilon_{w}(t)&\simeq &-\frac{{\rm Im}
\left(YY^\dagger\right)^2_{12}}{\left(YY^\dagger
\right)_{11}\left(YY^\dagger
\right)_{22}}
\frac{M_1}{M_2}\Gamma_{N_2}\frac{1}{(\Delta M)^2+ 
\Gamma_{N_2}^2}\nonumber\\
&\times& \left(2\,\Delta M \,{\rm sin}^2 \left[\frac{\Delta M t}{2}\right]
-\Gamma_{N_2}\,{\rm sin} \left[\Delta M t\right]\right),
\label{eq-08:ll}
\end{eqnarray}
where $\Delta M= (M_2-M_1)$. 
The CP asymmetry  (\ref{eq-08:ll}) is resonantly 
enhanced when  $\Delta M\simeq \Gamma_{N_2}$ and at the resonance
point it is given by

\begin{equation}
\label{eq-08:kkk}
\epsilon_{w}(t)\simeq -\frac{1}{2}\frac{{\rm Im}
\left(YY^\dagger\right)^2 _{12}}{\left(YY^\dagger
\right)_{11}\left(YY^\dagger
\right)_{22}}
\left(1-{\rm sin} \left[\Delta M t\right]-
{\rm cos} \left[\Delta M t\right]\right),
\end{equation}
The timescale for the 
building up of the CP asymmetry is $\sim 1/\Delta M$. 
The CP asymmetry  grows starting from a vanishing value and, for
$t\gg (\Delta M)^{-1}$, it averages to the constant 
standard value. 
This is true if the 
timescale for the other processes relevant
for leptogenesis is larger  than  $\sim 1/\Delta M$. 
In other words, one may define an ``average'' CP asymmetry

\begin{equation}
\langle \epsilon_{w}\rangle=\frac{1}{\tau_{\rm p}}
\int_{t-\tau_{\rm p}}^{t}
dt^\prime\, \epsilon^{{\rm W}}_{N_1}(t^\prime),
\end{equation}
where $\tau_{\rm p}$ represents the typical timescale 
of the other processes relevant for leptogenesis, {\it e.g.}
the $\Delta L=1$ scatterings. If $\tau_{\rm p}\gg 1/\Delta M\sim 
\Gamma^{-1}_{N_2}$, 
the oscillating functions
in (\ref{eq-08:kkk}) are averaged to zero and the average CP asymmetry
is given by the value used in the literature.  
However, the expression (\ref{eq-08:ll}) should be
used when   $\tau_{\rm p}\lsim 1/\Delta M\sim \Gamma^{-1}_{N_2}$. 

The fact that the CP asymmetry is a function of time is particularly
relevant in the case in which the asymmetry is generated by the decays
of two heavy states which are nearly degenerate in mass and oscillate
into one another with a timescale given by the inverse of the mass
difference and has been studied in Refs. 
\cite{DeSimone:2007pa,Garbrecht:2011aw}.  From eq. (\ref{eq-08:ll}) it is manifest
that the CP asymmetry itself oscillates with the very same timescale
and such a dependence may or may not be neglected depending upon the
rates of the other processes in the plasma.  If $\Gamma_{N_1}\gsim
\Gamma_{N_2}$, the time dependence of the CP asymmetry may not be
neglected.  The expression (\ref{eq-08:ll}) can also be used, once it is
divided by a factor 2 (because in the wave diagram also the charged
states of Higgs and lepton doublets may propagate) and the limit
$M_2\gg M_1$ is taken, for the CP asymmetry contribution from the
vertex diagram 
\begin{equation}
\epsilon_{v}(t)\simeq -\frac{{\rm Im}
\left(YY^\dagger\right)^2_{12}}{16 \pi\left(YY^\dagger
\right)_{11}}
\frac{M_1}{M_2}
\left(2\, {\rm sin}^2 \left[\frac{M_2 t}{2}\right]
-\frac{\Gamma_{N_2}}{M_2}\,{\rm sin} \left[M_2 t\right]\right).
\label{eq-08:lv}
\end{equation}
In this case, the timescale for this CP asymmetry is $\sim M_2$ and much larger
than any other timescale in the dynamics. Therefore, one can safely average 
over many oscillations, getting the expression present in the 
literature.

What discussed here provides only one example for which a quantum
Boltzmann approach is needed. In general, the lesson is that quantum
BE are relevant when the typical timescale in a
quantum physical process, such as the timescale for the
unflavour-to-flavour transition or the timescale to build up
coherently the CP asymmetry (of the order of $1/\Delta M$) is
larger than the timescale for the change of the abundances.

\vspace{1cm}
\section{Supersymmetric Leptogenesis}
\label{sec-08:susy}

\subsection{What's new?}
Supersymmetric leptogenesis constitutes a theoretically appealing
generalization of leptogenesis for the following reason: while the SM
equipped with the seesaw provides the simplest way to realize
leptogenesis, such a framework is plagued by an unpleasant fine-tuning
problem. For a non degenerate spectrum of heavy Majorana neutrinos,
successful leptogenesis requires generically a scale for the singlet
neutrino masses that is much larger than the EW
scale~\cite{Davidson:2002qv} but at the quantum level the gap between
these two scales becomes unstable.  Low-energy supersymmetry (SUSY)
can naturally stabilize the required hierarchy, and this provides a
sound motivation for studying leptogenesis in the framework of the
supersymmetrized version of the seesaw mechanism. Supersymmetric
leptogenesis, however, introduces a certain conflict between the
gravitino bound on the reheat temperature and the thermal production
of the heavy singlets neutrinos~\cite{Pagels:1981ke,Weinberg:1982zq,
  Khlopov:1984pf,Ellis:1984eq}. In this section, we will
not be concerned with the gravitino problem, nor with its possible
ways out but focus on the new features SUSY brings in for leptogenesis.

The supersymmetric type-I seesaw model is described by the superpotential
of the Minimal Supersymmetric SM (MSSM) with the additional terms:
\be
W=\frac{1}{2}M_{pq}N^c_{p}N^c_{q}+\lambda_{\alpha p}
N^c_{p}\,\ell_{\alpha} H_u,
\label{eq-08:superpotential}
\ee
where $p,q=1,2,\dots$ label the heavy singlet states in order of
increasing mass, and $\alpha=e,\mu,\tau$ labels the lepton flavour.
In \eqn{eq-08:superpotential} $\ell$, $H_u$ and $N^c$, are respectively
the chiral superfields for the lepton and the up-type Higgs $SU(2)_L$
doublets and for the heavy $SU(2)_L$ singlet neutrinos defined according
to usual conventions in terms of their LH Weyl spinor
components (for example the $N^c$ supermultiplet has scalar component
$\tilde N^*$ and fermion component $N^c_L$). Finally the $SU(2)_L$ index
contraction is defined as $\ell_{\alpha} H_u  
= \epsilon_{\rho\sigma}\ell_{\alpha}^\rho H_u^\sigma$ with $\epsilon_{12}=+1$.

Originally, the issue of MSSM leptogenesis was approached in
conjunction with SM leptogenesis~\cite{Covi:1996wh,Giudice:2003jh} as
well as in dedicated studies~\cite{Campbell:1992hd,Plumacher:1997ru}.
However in these first analysis, several features that are specific of
SUSY in the high temperature regime relevant for leptogenesis, in
which soft SUSY breaking parameters can be effectively set to zero,
had been overlooked. In that case, in spite of the large amount of new
reactions, the differences between SM and MSSM leptogenesis can be
resumed by means of simple counting of a few numerical
factors~\cite{DiBari:2004en,Strumia:2006qk,Davidson:2008bu}, like for
example the number of relativistic degrees of freedom in the thermal
bath, the number of loop diagrams contributing to the CP asymmetries,
the multiplicities of the final states in the decays of the heavy
neutrinos and sneutrinos and one can estimate~\cite{Davidson:2008bu}
\be
\frac{Y_{\Delta B}\left(\infty\right)^{\rm MSSM}}
  {Y_{\Delta
    B}\left(\infty\right)^{\rm SM}}
\approx 
\left\{
\begin{array}{ll}  
  \sqrt{2}  & \mbox{(strong washout)}; \\   
2\,\sqrt{2} & \mbox{(weak washout)}.
\end{array} 
\right. 
\ee

Recently it was pointed out in Ref.~\cite{Fong:2010qh} that in fact
MSSM leptogenesis is rich of new and non-trivial features, and
genuinely different from the simpler realization within the SM.
The two important effects are:\\
(a) If the SUSY breaking scale does not exceed by much $1\,$TeV, above
 $T\sim 5\times 10^7\,$GeV the particle and superparticle
density asymmetries do not equilibrate~\cite{Chung:2009qs}, 
and it is mandatory to account in the BE for the
differences in the number density asymmetries of the boson and fermion
degrees of freedom. This  can be given in terms of a
non-vanishing gaugino chemical potential.  \\
(b) When soft SUSY breaking parameters are neglected, additional
anomalous global symmetries that involve both $SU(2)_L$ and $SU(3)_C$
fermion representations emerge~\cite{Ibanez:1992aj}.  As a
consequence, the EW and QCD sphaleron equilibrium conditions are
modified with respect to the SM, and this yields a different pattern
of sphaleron induced lepton-flavour
mixing~\cite{Barbieri:1999ma,Nardi:2006fx,Abada:2006ea}.  In addition,
a new anomaly-free exactly conserved ${\cal R}$-symmetry provides an
additional constraint that is not present in the SM and a careful
counting reveals that \emph{four} independent quantities, rather than
the \emph{three} of the SM case, are required to give a complete
description of the various particle asymmetries in the thermal bath,
with the additional quantity corresponding to the number density
asymmetry of the heavy scalar neutrinos.

Although the modifications above are interesting from the theoretical
point of view, a quantitative comparison with the results obtained
when the new effects are ignored shows that the corrections 
remain below ${\cal
  O}(1)$~\cite{Fong:2010qh}\footnote{This modification would be
  important for certain supersymmetric leptogenesis scenarios which
  contain new sources of CP violation e.g. soft leptogenesis (see
  Section \ref{sec-08:soft}).}.  Finally, it should be pointed out
that in the supersymmetric case, the temperature in which the lepton
flavour effects (see Section \ref{sec-08:flavour}) come into play is
enhanced by a factor of $(1+\tan^2\beta)$ since the charged Yukawa
couplings are given by $h_\alpha = m_\alpha/(v_u
\cos\beta)$.

The purpose of the following sections is twofold. We describe the
chemical equilibrium conditions and conservation laws for MSSM in
conjunction to SM.  In Section~\ref{sec-08:general} we list the
constraints that hold independently of assuming a regime in which
particle and sparticle chemical potentials equilibrate
(superequilibration (SE) regime) or do not equilibrate
(non-superequilibration (NSE) regime).  In Section~\ref{sec-08:SE} we
list the constraints that hold only in the SE regime, and in
Section~\ref{sec-08:NSE} the ones that hold in NSE regime.  The
question of NSE is irrelevant in the SM since there are no
superparticles.  Hence the parts relevant for the SM are Section
\ref{sec-08:general} and Section \ref{sec-08:SE}, with the chemical
potential of the gaugino set to zero, the chemical potential of the
down-type higgsino replaced by the minus of the up-type higgsino (see
\eqns{eq-08:geq0}{eq-08:mueq0}), and all the quantities related to
superparticles replaced by the ones for particles.

\subsection{General constraints}
\label{sec-08:general}

We first list in items (1), (2) and (3) below the conditions that 
hold in the whole temperature range $M_W \ll T \lsim 10^{14}\,$GeV. 
Conversely, some of the Yukawa coupling conditions given in items (4) and (5)  
will have to be dropped as the temperature is increased and the corresponding
reactions go out of equilibrium.  
First we will relate the number density asymmetry of a particle $\Delta n \equiv n - \bar n$
for which a chemical potential can be defined to its chemical potential. 
For both bosons ($b$) and fermions ($f$) this relation acquires a particularly simple form
in the relativistic limit $m_{b,f}\ll T$, and at first order in the chemical potential
$\mu_{b,f}/T\ll 1$:
\be
\label{eq-08:Dnmu}
\Delta n_{b}=\frac{g_b}{3}
  T^2\mu_b, \qquad\quad
\Delta n_{f}=\frac{g_f}{6}T^2\mu_f\,.
\ee
For simplicity of notations, in the following we denote the chemical potentials with 
the same notation that labels the corresponding field: $\phi \equiv \mu_\phi$.


\begin{enumerate} \itemsep -1pt

\item[(1)] At scales much higher than $M_W$, gauge fields have vanishing
  chemical potential $W=B=g=0$~\cite{Harvey:1990qw}. This also implies that all
  the particles belonging to the same $SU(2)_L$ or $SU(3)_C$ multiplets 
  have the same chemical potential. For example
  $\phi(I_3=+\frac{1}{2})=\phi(I_3=-\frac{1}{2})$ for a field $\phi$
  that is a doublet of weak isospin $\vec I$, and similarly for color.

\item[(2)] Denoting by $\tilde W_R$, $\tilde B_R$ and $\tilde g_R$ the
  RH winos, binos and gluinos chemical potentials, and by
  $\ell,\,Q$ ($\tilde\ell,\,\tilde Q$) the chemical potentials of the
  (s)lepton and (s)quarks LH doublets, the following
  reactions: $\tilde Q +\tilde g_R \to Q$,\ $\tilde Q +\tilde W_R \to
  Q$,\ $\tilde \ell +\tilde W_R \to \ell $,\ $\tilde \ell +\tilde B_R
  \to \ell $,\ 
  imply that all gauginos have the same chemical potential:
%
$-\tilde g = Q-\tilde Q=
-\tilde W= \ell-\tilde \ell=-\tilde B,$
%
where $\tilde W$, $\tilde B$ and $\tilde g$
denote the chemical potentials of LH gauginos.
It follows that the chemical potentials of the SM particles are related 
to those of their respective superpartners as 
\
\begin{eqnarray}
  \label{eq-08:tQtell}
   \tilde{Q},\tilde \ell &=&    Q,\ell+  \tilde g \\
  \label{eq-08:HuHd}
   H_{u,d} &=&   \tilde H_{u,d}+  \tilde g \\
  \label{eq-08:tutdte}
   \tilde u,\tilde d,\tilde e  &=&   u,d,e-  \tilde g. 
\end{eqnarray}
The last relation, in which $u,d,e\equiv u_R,d_R,e_R$ denote the
RH $SU(2)_L$ singlets, follows e.g. from $ \tilde u^c_L= u^c_L+
\tilde g$ for the corresponding LH fields, together with
$u^c_L=-u_R$, and from the analogous relation for the $SU(2)_L$ singlet
squarks.

\item[(3)] Before EW symmetry breaking hypercharge is an exactly conserved
  quantity, and we can assume a vanishing  total hypercharge for the Universe: 
%
%
%
 \begin{equation}
    \label{eq-08:YtotY}
\sum_i\left(Y_{\Delta Q_i}+2Y_{\Delta u_i}-Y_{\Delta d_i}\right)
-\sum_\alpha\left(Y_{\Delta \ell_\alpha}+Y_{\Delta e_\alpha}\right)+
Y_{\Delta \tilde{H}_u}-Y_{\Delta \tilde{H}_d}= 0. 
\end{equation}

\item[(4)] When the reactions mediated by the lepton Yukawa couplings are
  faster than the Universe expansion rate\footnote{See Section \ref{sec-08:fla_temp} 
  for the temperature regime when lepton Yukawa interactions are in thermal equilibrium.}, 
  the following chemical equilibrium conditions are enforced:
\be 
\label{eq-08:leptons}
\ell_\alpha - e_\alpha + \tilde H_d + \tilde g =0, \qquad
(\alpha=e,\,\mu,\,\tau).  
\ee
%
%
%
If the temperature is not too low lepton flavour equilibration (see Section \ref{sec-08:LFE}) 
induced by off-diagonal slepton soft masses will not occur. We assume that this
is the case, and thus we take the three $\ell_\alpha$ to be 
independent quantities.

\item[(5)]  Reactions mediated by the quark Yukawa couplings enforce the
  following six  chemical equilibrium conditions:
\bea 
\label{eq-08:upquarks}
Q_i - u_i + \tilde H_u + \tilde g &=&0, \qquad
(u_i=u,\,c,\,t),\\  
\label{eq-08:downquarks}
Q_i - d_i + \tilde H_d + \tilde g &=&0, \qquad
(d_i=d,\,s,\,b)\,.  
\eea
The up-quark Yukawa coupling maintains chemical equilibrium between
the LH and RH up-type quarks up to $T\sim 2\cdot
10^6\,$GeV.  Note that when the Yukawa reactions of at least two
families of quarks are in equilibrium, the mass basis is fixed for all
the quarks and squarks.  Intergeneration mixing then implies that
family-changing charged-current transitions are also in equilibrium:
$b_L \to c_L$ and $t_L \to s_L$ imply $Q_2 = Q_3$; $s_L \to u_L$ and
$c_L \to d_L$ imply $Q_1 = Q_2$. Thus, up to temperatures $T\lsim
10^{11}\,$GeV, that are of the order of the equilibration temperature
for the charm Yukawa coupling, the three quark doublets have the same
chemical potential:
  \begin{equation}
    \label{eq-08:Q}
    Q\equiv Q_3=Q_2=Q_1. 
  \end{equation}
  At higher temperatures, when only the third family is in
  equilibrium, we have instead $Q\equiv Q_3=Q_2\neq Q_1$.  Above
  $T\sim 10^{13}$ when (for moderate values of $\tan\beta$) 
  also $b$-quark (as well as the $\tau$-lepton) 
  $SU(2)_L$ singlets decouple from their Yukawa
  reactions, all intergeneration mixing becomes negligible and
  $Q_3\neq Q_2\neq Q_1$.

\end{enumerate}

\subsection{Superequilibration regime}
\label{sec-08:SE}

At relatively low temperatures, additional conditions from reactions
in chemical equilibrium hold. Since the constraints below apply
only in the SE regime, we number them including this label.
\begin{enumerate}  \itemsep -1pt
\item[$6_{\rm SE}$.] 
  Equilibration of the particle-sparticle chemical potentials
  $\mu_\phi=\mu_{\tilde \phi}$~\cite{Chung:2009qs} is ensured when reactions
  like $\tilde \ell\tilde \ell \to \ell\ell$ are faster than the
  Universe expansion rate.  These reactions are induced by gaugino
  interactions, but since they require a gaugino chirality flip they
  turn out to be proportional to its mass $m_{\tilde g}$, and 
  can be neglected in the limit  $m_{\tilde g}\to 0$.

  Furthermore, since the $\mu$ parameter of the $H_u H_d$
  superpotential term is expected to be of the order of the soft
  gaugino masses, it is reasonable to consider in the same temperature
  range also the effect of the higgsino mixing term, which implies
  that the sum of the up- and down- higgsino chemical potentials
  vanishes.  The rates of the corresponding reactions, given
  approximately by $\Gamma_{\tilde g} \sim m^2_{\tilde g}/T$ and
  $\Gamma_{\mu} \sim \mu^2/T$, are faster than the Universe expansion
  rate up to temperatures
\be
\label{eq-08:Tgmu}
 T\lsim 5\cdot 10^7 
\left(\frac{m_{\tilde g},\,\mu}{500\,{\rm GeV}}\right)^{2/3}\, {\rm GeV}.
\ee
The corresponding chemical equilibrium relations enforce the
conditions:
\bea
\label{eq-08:geq0}
\tilde g &=&0, \\ 
\label{eq-08:mueq0}
\tilde H_u+\tilde H_d&=&0.  
\eea
\item[$7_{\rm SE}$.]  Up to temperatures given by \eqn{eq-08:Tgmu} the
  MSSM has the same global anomalies than the SM, that are the EW
  $SU(2)_L$-$U(1)_{B+L}$ mixed anomaly and the QCD chiral anomaly. They
  generate the effective operators $O_{EW}=\Pi_\alpha(QQQ\ell_\alpha)$~\cite{Bento:2003jv}
  and $O_{QCD}=\Pi_i(QQu^c_{Li}d^c_{Li})$~\cite{Mohapatra:1991bz,Moore:1997im,Bento:2003jv}. 
  Above the EW phase transition reactions induced by these operators are in thermal
  equilibrium, and the corresponding conditions read:
\bea
\label{eq-08:EW}
&&9\,Q+\sum_\alpha \ell_\alpha = 0 \\
\label{eq-08:QCD}
&&6\,Q-\sum_i\left(u_i+d_i\right)=0\,,
\eea
where we have used the same chemical potential for the three quark doublets
(\eqn{eq-08:Q}), which is always appropriate in the SE regime below the limit
\eqn{eq-08:Tgmu}.

\end{enumerate}

Eqs.~(\ref{eq-08:leptons}) and (\ref{eq-08:upquarks})--(\ref{eq-08:Q}), 
together with the SE conditions eqs.~(\ref{eq-08:geq0})--(\ref{eq-08:mueq0}), the two
anomaly conditions eqs.~(\ref{eq-08:EW})--(\ref{eq-08:QCD}) and the hypercharge
neutrality condition eq.~(\ref{eq-08:YtotY}), give $11+2+2+1=16$ constraints
for the 18 chemical potentials.  Note however that there is one
redundant constraint, that we take to be the QCD sphaleron condition,
since by summing up eqs.~(\ref{eq-08:upquarks}) and (\ref{eq-08:downquarks})
and taking into account eqs.~(\ref{eq-08:Q}), (\ref{eq-08:geq0}), and
(\ref{eq-08:mueq0}) we obtain precisely \eqn{eq-08:QCD}. Therefore,
like in the SM, we have three independent chemical potentials. 
We can define three linear combinations of the chemical potentials
corresponding to the $SU(2)_L$ anomaly-free flavour charges
$\Delta_\alpha\equiv B/3-L_\alpha$ that being anomaly-free 
and perturbatively conserved by the low energy MSSM
Lagrangian, evolve {\it slowly} because the corresponding symmetries
are violated only by the heavy Majorana neutrino dynamics. 
Their evolution needs to be computed by means of three independent BE. 
In terms of the abundances \eqn{eq-08:Y} the density of 
the $\Delta_\alpha$ charges normalized to the entropy density can be written as:
\be 
\label{eq-08:YDeltaAlpha}
Y_{\Delta_\alpha} = 3\,\left[\frac{1}{3}
    \sum_i\left(2Y_{\Delta Q_i}+Y_{\Delta u_i}+Y_{\Delta d_i}\right)-
  (2Y_{\Delta \ell_\alpha}+Y_{\Delta
    e_\alpha})-\frac{2}{3}Y_{\Delta\tilde g}\right]\,. 
\ee
The expression above is completely general and holds in all
temperature regimes, including the NSE regime (see Section \ref{sec-08:NSE}).  

The density asymmetries of the doublet leptons and higgsinos, that
weight the washout terms in the BE, can now be
expressed in terms of the anomaly-free charges by means of the $A$
matrix and $C$ vectors introduced respectively in Ref.~\cite{Barbieri:1999ma}
and Ref.~\cite{Nardi:2006fx} that are defined as:
\be
\label{eq:AC}
Y_{\Delta\ell_\alpha}= A^\ell_{\alpha\beta}\,\, Y_{\Delta_\beta}, 
\qquad\qquad 
Y_{\Delta \tilde H_{u,d}}= 
C^{\tilde H_{u,d}}_\alpha\,\,  Y_{\Delta_\alpha}. 
\ee
Here and in the following we will give results for the $A$ and $C$
matrices for the fermion states.  We recall that in the SE regime the
density asymmetry of a scalar boson that is in chemical equilibrium with
its fermionic partner is given simply by $Y_{\Delta b}=2\,Y_{\Delta f}$
with the factor of 2 from statistics.

\subsubsection{First generation Yukawa reactions out of equilibrium
  (SE regime)}
\label{sec-08:ACT3SE}

As an example let us now consider the temperatures $T\gsim 4\cdot
10^6(1+\tan^2\beta)\,{\rm GeV,} $ when the $d$-quark Yukawa coupling
can be set to zero (in order to remain within the SE regime we assume
$\tan\beta \sim 1$).  In this case the equilibrium dynamics is
symmetric under the exchange $u \leftrightarrow d$ (both chemical
potentials enter only the QCD sphaleron condition \eqn{eq-08:QCD} with
equal weights) and so must be any physical solution of the set of
constraints. Thus, the first condition in \eqn{eq-08:downquarks} can be
replaced by the condition $d=u$, and again three independent
quantities suffice to determine all the particle density asymmetries.
The corresponding result is:
\begin{eqnarray}
A^\ell=\frac{1}{3\times 2148}
\left(\begin{array}{ccc}
-906 &\ \,\, 120  &\ \,\, 120\\
\ \, 75 &-688 &\ \, 28\\
\ \, 75 &\ \,28  &-688 
\end{array}\right), &\;& C^{\tilde H_u}=-C^{\tilde H_d}=
\frac{-1}{2148}\left(37,\;52,\;52\right). \qquad
\label{eq-08:ACT3SE}
\end{eqnarray}
 Note that since in this regime the chemical potentials for the scalar
and fermion degrees of freedom of each chiral multiplet equilibrate,
the analogous results for $Y_{\Delta\ell_\alpha}+Y_{\Delta\tilde\ell_\alpha}$ 
can be obtained by simply multiplying the $A$ matrix in \eqn{eq-08:ACT3SE} by a factor of 3. 
This gives the same $A$ matrix obtained in the non-supersymmetric
case in the same regime (see e.g. eq.~(4.12) in Ref.~\cite{Nardi:2006fx}).
The $C$ matrix (multiplied by the same factor of 3) differs from the
non-supersymmetric result by a factor $1/2$. This is because after
substituting $\tilde H_d=-\tilde H_u$ (see \eqn{eq-08:mueq0}) all
the chemical potential conditions are formally the same than in the SM
with $\tilde H_u$ identified with the chemical potential of the scalar
Higgs, but since $C$ expresses the result for number densities, in the
SM a factor of 2 from boson statistics appears for the SM Higgs.  This
agrees with the analysis in Ref.~\cite{Inui:1993wv}, and is a general result
that holds for SUSY within the SE regime.

\subsection{Non-superequilibration regime}
\label{sec-08:NSE}

At temperatures above the limit given in \eqn{eq-08:Tgmu} the Universe
expansion is fast enough that reactions induced by $m_{\tilde g}$ and
$\mu$ do not occur. Setting to zero in the high temperature effective
theory these two parameters has the following consequences:
\begin{itemize} \itemsep -1pt
\item Condition \eqn{eq-08:geq0} has to be dropped, and gauginos
  acquire a non-vanishing chemical potential $\tilde g\neq 0$
  (corresponding to the difference between the number of LH and RH
  helicity states). The chemical potentials of the members of the same
  matter supermultiplets are no more equal (non-superequilibration)
  but related as in eqs.~(\ref{eq-08:tQtell})--(\ref{eq-08:tutdte}).

\item Condition \eqn{eq-08:mueq0} also has to be dropped, and the
  chemical potentials of the up- and down-type Higgs and higgsinos do
  not necessarily sum up to zero.

\item The MSSM gains two new global symmetries: $m_{\tilde
    g}\to 0$ yields a global $R$-symmetry, while $\mu\to 0$
  corresponds to a  global symmetry of the Peccei-Quinn ($PQ$)
  type.
\end{itemize}

\subsubsection{Anomalous and non-anomalous symmetries}
\label{sec-08:symmetries}

Two linear combinations $R_2$ and $R_3$ of $R$ and $PQ$, 
having respectively only $SU(2)_L$ and $SU(3)_C$ mixed
anomalies have been identified in Ref.~\cite{Ibanez:1992aj}
\be
\label{eq-08:R2_R3}
R_2 = R-2\,PQ, \;\;\;\;\;\;\;\;
R_3 = R-3\,PQ\,. 
\ee
The authors of Ref.~\cite{Ibanez:1992aj} have also constructed the effective 
multi-fermions operators generated by the
mixed anomalies:
\bea
\label{eq-08:tO-EW}
\tilde O_{EW} &=&
\Pi_\alpha \left(QQQ\ell_\alpha\right)\; \tilde H_u\tilde H_d\;\tilde W^4\,,\\
\label{eq-08:tO-QCD}
\tilde O_{QCD} &=& \Pi_i \left(QQu^c d^c\right)_{i}\; \tilde g^6 \,.
\eea

Given that three global symmetries $B$, $L$ and $R_2$ have mixed
$SU(2)_L$ anomalies (but are free of $SU(3)_C$ anomalies) 
we can construct two anomaly-free combinations, the first one being 
$B-L$ which is only violated perturbatively by $N^c\,\ell H_u$ and 
the second anomaly-free combination which is also an exact symmetry 
of the MSSM+seesaw in the NSE regime~\cite{Fong:2010qh}
\be
\label{eq-08:calR}
 {\cal R}=\frac{5}{3}B-L+R_2,  
\ee
and is exactly conserved.  In the $SU(3)_C$ sector, besides the chiral
anomaly we now have also $R_3$ mixed anomalies. Thus also in this
case anomaly-free combinations can be constructed, and  in particular we
can define one combination for each quark superfield.
Assigning to the LH supermultiplets chiral
charge $\chi=-1$ these combinations have the form~\cite{Fong:2010qh}:
\be
\label{eq-08:chiral} 
\chi_{q_L}+\kappa_{q_L}\,R_3,
\ee
where, for example, $\kappa_{u^c_L}=\kappa_{d^c_L}=1/3$ and
$\kappa_{Q_L}=2/3$.  Note that since $R_3$ is perturbatively conserved
by the complete MSSM+seesaw Lagrangian, when the Yukawa coupling of
one quark is set to zero the corresponding charge \eqn{eq-08:chiral}
will be exactly conserved.

\subsubsection{Constraints in the non-superequilibration regime}
\label{sec-08:NSEconstraints}

In the NSE regime, the conditions listed in items
$6_{SE}$ and $7_{SE}$ of the previous section have to be dropped, but
new conditions arise.

\begin{enumerate} \itemsep -1pt

\item[$6_{NSE}$.]  The conservation law for the 
${\cal R}$ charge yields the following global neutrality condition:
 \begin{eqnarray}
   \nonumber
 {\cal R}_{\rm tot} &=&  
\sum_f \Delta n_f  {\cal R}_f + 
\sum_b \Delta n_b  {\cal R}_b +
\Delta n_{\tilde N_1}  {\cal R}_{\tilde N_1}\\
 &=& \frac{T^2}{6}\left( \sum_i\left(2Q_i-5u_i+4d_i\right)+ 
2\sum_\alpha\left(\ell_\alpha+e_\alpha\right)+5\tilde H_d-\tilde H_u
+31\,\tilde g \right) 
-\Delta n_{\tilde N_1}
=0.\qquad \quad
    \label{eq-08:calRtot}
  \end{eqnarray}
The last terms in both lines of \eqn{eq-08:calRtot} correspond to the
contribution to ${\cal R}$-neutrality from the lightest sneutrino
asymmetry $\Delta n_{\tilde N_1} = n_{\tilde N_1}- n_{\tilde N_1^*}$
with charge ${\cal R}_{\tilde N_1}= - {\cal R}_{N^c}=-1$.  
Note that since in general $\tilde N_1$ is not in chemical equilibrium, no
chemical potential can be associated to it, and hence this constraint
needs to be formulated in terms of its number density asymmetry that
has to be evaluated by solving a BE for
$Y_{\Delta_{\tilde N}} \equiv Y_{\tilde N_1}- Y_{\tilde N_1^*}$ (see
Section \ref{sec-08:MSSM_BE}).  

\item[$7_{NSE}$.] The operators in eqs.~(\ref{eq-08:tO-EW})--(\ref{eq-08:tO-QCD}) 
induce transitions that in the NSE regime are in chemical equilibrium. 
This enforces the generalized EW and QCD sphaleron equilibrium
conditions~\cite{Ibanez:1992aj}: 
\bea
\label{eq-08:tEWmu}
&&3\sum_i Q_i+\sum_\alpha\ell_\alpha
+\tilde H_u+\tilde H_d+4\,\tilde g=0,\\
\label{eq-08:tQCDmu}
&&2\sum_i Q_i-\sum_i\left(u_i+d_i\right)+6\,\tilde g=0, 
\eea
that replace eqs.~(\ref{eq-08:EW}) and (\ref{eq-08:QCD}).

\item[$8_{NSE}$.] The chiral-$R_3$ charges in \eqn{eq-08:chiral} are
  anomaly-free, but clearly they are not conserved by the quark
  Yukawa interactions. However, when a quark supermultiplet decouples
  from its Yukawa interactions an exact conservation law arises\footnote{Note
  that $h_{u,d} \to 0$ implies $u$ and $d$ decoupling, but $Q_1$
  decoupling is ensured only if also $h_{c,s}\to 0$.}.  
  The conservation laws corresponding to these symmetries read:
\bea
\label{eq-08:chiralmuquarks}
\frac{T^2}{6}\left[3 q_R + 6(q_R-\tilde g)\right]+\frac{1}{3}
\,R_{3\;\rm tot}
&=&0 \\
\label{eq-08:chiralmuquarks2}
\frac{T^2}{6}\,2\,\left[3 Q_L + 6(Q_L+\tilde g)\right]-\frac{2}{3}
\,R_{3\;\rm tot}
&=&0
\eea  
and hold for $q_R=u_i,\,d_i$ and $Q_L=Q_i$ in the regimes when the
appropriate Yukawa reactions are negligible.  Note the factor of 2 for
the $Q_L$ chiral charge in front of the first square bracket in
\eqn{eq-08:chiralmuquarks2} that is due to $SU(2)_L$ gauge multiplicity.
In terms of chemical potentials and $\Delta n_{\tilde N_1}$, 
the total $R_3$ charge in \eqns{eq-08:chiralmuquarks}{eq-08:chiralmuquarks2} reads:
\bea
\label{eq-08:R3tot}
R_{3\;\rm tot} & =&  \frac{T^2}{6\,}
\left(82\,\tilde g -3\sum_i \left(2\,Q_i+11\,u_i-4\,d_i\right)
+ 
\sum_\alpha\left(16\,\ell_\alpha+13\,e_\alpha\right)
+16\,\tilde H_d-14\,\tilde H_u.\right) 
 \nonumber \\  & & 
+\Delta n_{\tilde N_1} {R_3}_{\tilde N_1},
\eea
where ${R_3}_{\tilde N_1} =-1$. 
As regards the leptons, since they do not couple to the QCD anomaly,
by setting $h_{e}\to 0$ a symmetry under chiral supermultiplet
rotations is directly gained for the RH leptons implying $\Delta
n_e+\Delta {\tilde n_e}=0$ and giving the condition:
\be
\label{eq-08:chiralmuleptons}
e-\frac{2}{3}\tilde g =0.
\ee
No analogous condition arises for the lepton doublets relevant for
leptogenesis, since by assumption they remain coupled via Yukawa
couplings to the heavy $N$'s.
\end{enumerate}

In the NSE regime there are different flavour mixing matrices for the
scalar and fermion components of the leptons and Higgs
supermultiplets.  To express more concisely all the results, it is
convenient to introduce a new $C$ vector to describe the gaugino
number density asymmetry per degree of freedom in terms of the
relevant charges:
\bea
\label{eq:Cg}
Y_{\Delta \tilde g}=C_a^{\tilde g}\;Y_{\Delta_a}\, ,  
&\;\;\;\;\;\;\;{\rm with}\;\;\;\;\;\;\;\;\;
&\Delta_{a}=\left(\Delta_\alpha,\Delta_{\tilde N}\right)\; .
\eea

\subsubsection{First generation Yukawa reactions out of equilibrium
  (NSE regime)} 
\label{sec-08:ACT3NSE}

As an example, in the temperature range between $10^8$ and $10^{11}\,$GeV, and for
moderate values of $\tan\beta$, all the first generation Yukawa
couplings can be set to zero.  Using for $u,\,d$ 
conditions \eqn{eq-08:chiralmuquarks} and for $e$ 
condition \eqn{eq-08:chiralmuleptons} as are implied  
by $h_{u,d},\,h_e\to 0$ we obtain:  
\begin{eqnarray}
\nonumber
A^\ell&=&\frac{1}{9\times 162332}\left(
\begin{array}{cccc}
-198117 &\ \,\, 33987  &\ \,\, 33987& -8253 \\
\ \, 26634 &-147571  &\ \, 14761&-8055\\
\ \, 26634 &\ \, 14761   &-147571&-8055
\end{array}
\right), \\ [5pt]
\nonumber
C^{\tilde g}&=&\frac{-11}{162332}\left(163,\;
165,\;165,\;-255\right),\quad \\ [5pt]
\nonumber
C^{\tilde H_u}&=& 
\frac{-1}{162332}\left(3918,\;
4713,\; 4713,\; 95 \right)\,, \\ [5pt]
C^{\tilde H_d} &=& \frac{1}{3\times 162332  }\left(5413,\;
9712,\;9712,\;-252\right)\,,  
\label{eq-08:ACT3NSE}
\end{eqnarray}
where the rows correspond to $( Y_{\Delta_e},\; Y_{\Delta_\mu},\;
Y_{\Delta_\tau},\; Y_{\Delta_{\tilde N}})$.  For completeness, in
\eqn{eq-08:ACT3NSE} we have also given the results for $C^{\tilde H_d}$
even if only the up-type Higgs density asymmetry is relevant for the
leptogenesis processes.  Note that neglecting the contribution of
$\Delta n_{\tilde N_1}$ to the global charges ${\cal R}_{\rm tot}$ in
\eqn{eq-08:calRtot} and ${R_3}_{\rm tot}$ in \eqn{eq-08:R3tot} corresponds
precisely to setting to zero the fourth column in all the previous
matrices. Then, analogously with the SE and SM cases, within this
`3-column approximation' all particle density asymmetries can be
expressed just in terms of the three $Y_{\Delta_\alpha}$ charge
densities.

\subsection{Supersymmetric Boltzmann equations}
\label{sec-08:MSSM_BE}

In order to illustrate how the new effects described above
modify the structure of the BE, here we write down a simpler
expressions in which only decays and inverse decays are included\footnote{The 
complete set of BE including decays, inverse decays 
and scatterings with top-quark is given in the Appendix of Ref.~\cite{Fong:2010qh}.}:
\begin{eqnarray}
sHz\frac{d{Y}_{N_1}}{dz} & = & -\left(\frac{Y_{N_1}}{Y_{N_1}^{eq}}-1\right)\gamma_{N_1},
\label{eq-08:YN}\\
sHz\frac{d{Y}_{\tilde N_+}}{dz} & = & 
-\left(\frac{Y_{\tilde N_+}}{Y_{\tilde{N}_1}^{eq}}-2\right)
\gamma_{\tilde N_1},
\label{eq-08:YNplus}\\
sHz\frac{d{Y}_{\Delta_{\tilde N}}}{dz} & = & 
-\frac{Y_{\Delta_{\tilde N}}}{Y_{\tilde{N}_1}^{eq}}\gamma_{\tilde N_1}
-\frac{3}{2}\,\gamma_{\tilde N_1}
\sum_a C_a^{\tilde g}\;\frac{Y_{\Delta_a}}{Y^{eq}_\ell}
 +\dots\,,
 \label{eq-08:YNminus}\\
sHz\frac{d{Y}_{\Delta_{\alpha}}}{dz}  &=&  -\epsilon_{\alpha}
\left[\left(\frac{Y_{N_1}}{Y_{N_1}^{eq}}-1\right)
\gamma_{N_1}+\left(\frac{Y_{\tilde N_+}}{Y_{\tilde{N}_1}^{eq}}-2\right)
\gamma_{\tilde N_1}\right] 
 \nonumber  \\ &&
+ P_{1\alpha}^0\left(\gamma_{N_1}+\frac{1}{2}\gamma_{\tilde N_1}\right)
\sum_a\left(A_{\alpha a}^\ell+ C_a^{\tilde H_u} 
+ C_a^{\tilde g}\right)\frac{Y_{\Delta_a}}{Y^{eq}_\ell}\,,
\label{eq-08:YDelta}
\end{eqnarray}
where $\gamma_{\tilde N_1}$ is the corresponding thermally 
averaged decay rate for RH sneutrino $\tilde N_1$. 
In \eqn{eq-08:YNplus} we have introduced the overall 
sneutrino abundance $Y_{\tilde N_+}=Y_{\tilde
  N_1}+ Y_{\tilde N_1^*}$, while $Y_{\Delta_{\tilde N}} \equiv Y_{\tilde
  N_1}- Y_{\tilde N_1^*}$ in \eqn{eq-08:YNminus} is the sneutrino density
asymmetry that was already introduced in Section
\ref{sec-08:NSEconstraints}.  In the washout terms we have normalized the
charge densities $Y_{\Delta_a}=(Y_{\Delta_\alpha},\;Y_{\Delta_{\tilde{N}}}) $ 
to the equilibrium density of a fermion with one degree of freedom $Y_\ell^{eq}$. 
In \eqns{eq-08:YN}{eq-08:YDelta} we have also
neglected for simplicity all finite temperature effects.  Taking these
effects into account would imply for example that the CP asymmetry for
$\tilde N$ decays into fermions is different from the one for decays
into scalars~\cite{Giudice:2003jh}, while we describe both CP asymmetries 
with $\epsilon_\alpha$. A few remarks regarding \eqn{eq-08:YNminus} are in
order.  In the SE regime $\tilde g=0$ and thus it would seem that the
sneutrino density asymmetry $Y_{\Delta_{\tilde N}}$ vanishes.
However, this only happens for decays and inverse decays, and it is no
more true when additional terms related to scattering processes, that
are represented in the equation by the dots, are also included (see
Ref.~\cite{Plumacher:1997ru} and the Appendix of Ref.~\cite{Fong:2010qh}).  
Therefore, also in the SE regime $Y_{\tilde N_1}$ and $Y_{\tilde N_1^*}$ in general
differ. However, in this case recasting their equations in terms of
two equations for $Y_{\tilde N_+}$ and $Y_{\Delta_{\tilde N}}$ is just
a convenient parametrization. On the contrary, in the NSE regime this
is mandatory, because the sneutrinos carry a globally conserved
${\cal R}$-charge and $Y_{\Delta_{\tilde N}}$ is required to formulate
properly the corresponding conservation law. As we have seen, this
eventually results in $Y_{\Delta_{\tilde N}}$ contributing to the
expressions of the lepton flavour density asymmetries in terms of
slowly varying quantities.

In Ref.~\cite{Fong:2010qh} a complete numerical analysis was carried
out and it was shown that numerical corrections with respect to
the case when NSE effects are neglected remain at the ${\cal O}(1)$
level. This is because only spectator processes get affected, while
the overall amount of CP violation driving leptogenesis remains the
same than in previous treatments.

\vspace{1cm}
\section{Beyond Type-I Seesaw and Beyond the Seesaw}
\label{sec-08:beyond}

There exist many variants of leptogenesis models beyond the standard
type-I seesaw.  In this section, we try to classify them into
appropriate groups.  Unavoidably there would be some overlap i.e. a
hybrid model of leptogenesis which can belong to more than one group,
e.g. soft leptogenesis (Section \ref{sec-08:soft}) from resonantly
enhanced CP asymmetry could rightly fall under resonant leptogenesis
(Section \ref{sec-08:resonant}). However we try our best to categorize
them according to the main features of the model and, when
appropriate, they will be quoted in more than one place.  Clearly, the
number of beyond type-I seesaw leptogenesis models is quite large. We
have not attempted in any way to be exhaustive, and we apologize in
advance for the unavoidable several omissions.

\subsection{Resonant leptogenesis}
\label{sec-08:resonant}
A resonant enhancement of the CP asymmetry in $N_1$ decay occurs when
the mass difference between $N_1$ and $N_2$ is of the order of the
decay widths. Such a scenario has been termed `resonant leptogenesis',
and has benefited from many studies in different formalisms\footnote{See 
Ref.~\cite{Rangarajan:1999kt} for a comparison of the
  different calculations.}
\cite{Flanz:1994yx,Covi:1996fm,Pilaftsis:1997jf,Pilaftsis:2003gt,Hambye:2004jf,
  Pilaftsis:2004xx,Albright:2004ws,Albright:2005bm,Pilaftsis:2005rv,Anisimov:2005hr,
  Xing:2006ms,West:2006fs,Cirigliano:2006nu,Branco:2006hz,DeSimone:2007rw,DeSimone:2007pa,
  Babu:2007zm,Cirigliano:2007hb,Pilaftsis:2008qt,Deppisch:2010fr,Iso:2010mv,Gu:2010ye}
(see Ref.~\cite{Pilaftsis:2009pk} for a recent review).    The
resonant effect is related to the self energy contribution to the CP
asymmetry. Consider, for simplicity, the case where only $N_2$ is
quasi-degenerate with $N_1$. Then, the self-energy contribution
involving the intermediate $N_2$, to the total CP asymmetry (we
neglect important flavour effects \cite {Pilaftsis:2005rv}) is given
by \be \epsilon_{1}({\rm
  self-energy})=-\frac{M_1}{M_2}\frac{\Gamma_{N_2}}{M_2} \frac{M_2^2
  (M_2^2 - M_1^2)}{(M_2^2 - M_1^2)^2+M_1^2\Gamma_{N_2}^2} \frac{{\rm
    Im}[(\lambda^\dagger\lambda)_{12}^2]}
{(\lambda^\dagger\lambda)_{11}(\lambda^\dagger\lambda)_{22}}.  \ee The
resonance condition reads \be |M_2-M_1| =\frac{\Gamma_{N_2}}{2}.  \ee
In this case \be |\epsilon_{1}({\rm resonance})|\simeq\frac{1}{2}
\frac{|{\cal I}m[(\lambda^\dagger\lambda)_{12}^2]|}
{(\lambda^\dagger\lambda)_{11}(\lambda^\dagger\lambda)_{22}}.  \ee
Thus, in the resonant case, the asymmetry is suppressed by neither the
smallness of the light neutrino masses, nor the smallness of their
mass splitting, nor small ratios between the singlet neutrino
masses. Actually, the CP asymmetry could be of order one (more
accurately, $|\epsilon_1|\leq1/2$).

With resonant leptogenesis, the BE are different.
The densities of $N_1$ and $N_2$ are followed, since both
contribute to the asymmetry, and the relevant timescales are
different.  For instance, the typical time-scale to build up
coherently the CP asymmetry is unusually long, of order $1/\Delta
M$. In particular, it can be larger than the time-scale for the change
of the abundance of the sterile neutrinos. This situation implies that
for resonant leptogenesis quantum effects in the BE can be
significant~\cite{DeSimone:2007rw,DeSimone:2007pa,Garny:2011hg,Garbrecht:2011aw} 
(see Section \ref{sec-08:quantum}).

The fact that the asymmetry could be large, independently of the
singlet neutrino masses, opens up the possibility of low scale
resonant leptogenesis. Models along these lines have been constructed
in Refs. \cite{Hambye:2004jf,Pilaftsis:2005rv,West:2006fs,West:2004me}.
It is  a  theoretical challenge to construct models
where a mass splitting as small as the decay width is naturally
achieved. For attempts that utilize approximate flavour symmetries see,
for example, Refs.~\cite{Pilaftsis:2004xx,Albright:2004ws,Xing:2006ms,
Babu:2007zm,Branco:2005ye,Deppisch:2010fr},
while studies of this issue in the framework of minimal flavour violation can
be found in Refs.~\cite{Cirigliano:2006nu,Branco:2006hz,Cirigliano:2007hb}.
The possibility of observing resonant CP violation due to heavy RH neutrinos
at the LHC was studied in Refs.~\cite{Bray:2007ru,Blanchet:2009bu}.

\subsection{Soft leptogenesis}
\label{sec-08:soft}
The modifications to standard type-I leptogenesis due to SUSY have
been discussed in Section \ref{sec-08:susy}. The important parameters
there are the Yukawa couplings and the singlet neutrino parameters,
which appear in the superpotential \eqn{eq-08:superpotential}.  SUSY
must, however, be broken. In the framework of the MSSM extended to
include singlet neutrinos (MSSM+N), there are, in addition to the soft
SUSY breaking terms of the MSSM, terms that involve the singlet
sneutrinos $\widetilde N_i$, in particular bilinear ($B$) and
trilinear ($A$) scalar couplings. These terms provide additional
sources of lepton number violation and of CP violation.  Scenarios
where these terms play a dominant role in leptogenesis have been
termed `soft
leptogenesis'~\cite{Boubekeur:2002jn,Grossman:2003jv,D'Ambrosio:2003wy,Chun:2004eq,
  Boubekeur:2004ez,D'Ambrosio:2004fz,Chen:2004xy,Kashti:2004vj,Grossman:2005yi,
  Ellis:2005uk,Chun:2005ms,Medina:2006hi,Garayoa:2006xs,Chun:2007ny,BahatTreidel:2007ic,
  Fong:2008mu,Fong:2008yv,Fong:2009iu,Babu:2009pi,Kajiyama:2009ae,Garayoa:2009my,
  Fong:2010zu,Hamaguchi:2010cw,Fong:2010bv} (see also 
Ref.~\cite{Fong:2011yx} for a recent review).

Soft leptogenesis can take place even with a single RH neutrino
because the presence of the $B$ term implies that $\widetilde N$ and
$\widetilde N^\dagger$ states mix to form two mass eigenstates with
mass splitting proportional to $B$ itself. Furthermore when $B \sim
\Gamma_{\widetilde N}$, the CP asymmetry is resonantly enhanced
realizing the resonant leptogenesis scenario (see Section
\ref{sec-08:resonant}).  In the following we will consider a single
generation MSSM+N.  The relevant soft SUSY terms involving
$\widetilde{N}$, the $SU(2)_L$ gauginos
$\widetilde{\lambda}_{2}^{\pm,0}$, the $U(1)_Y$ gauginos
$\widetilde{\lambda}_{1}$ and the three sleptons
$\widetilde{\ell}_\alpha$ in the basis in which charged lepton
Yukawa couplings are diagonal are given by
\bea
\! -\mathcal{L}_{\rm soft} 
& = & \widetilde{M}^{2}\widetilde{N}^{*}\widetilde{N}
+\left(A Y_{\alpha}\epsilon_{ab}\widetilde{N}
\widetilde{\ell}_{\alpha}^{a}H_{u}^{b}+\frac{1}{2}BM
\widetilde{N}\widetilde{N}+{\rm h.c.}\right)
\nonumber \\ 
&  & +\frac{1}{2}\left(m_{2}\overline{\widetilde{\lambda}_{2}^{\pm,0}}
P_{L}\widetilde{\lambda}_{2}^{\pm,0}
+m_{1}\overline{\widetilde{\lambda}_{1}}P_{L}\widetilde{\lambda}_{1}
+{\rm h.c.}\right), 
\label{eq-08:soft_terms}
\eea
where for simplicity, proportionality of the bilinear and trilinear
soft breaking terms to the corresponding SUSY invariant
couplings has been assumed: $B_M=BM$ and $A_{\alpha}=AY_{\alpha}$.
The Lagrangian derived from  \eqns{eq-08:superpotential}{eq-08:soft_terms} 
is characterized by only three independent physical phases:
$\phi_{A} \equiv \arg\left(AB^{*}\right)$, 
$\phi_{g_{2}} \equiv \frac{1}{2}\arg\left(Bm_{2}^{*}\right)$
and $\phi_{g_{Y}} \equiv \frac{1}{2}\arg\left(Bm_{1}^{*}\right)$
which can be assigned to $A$, and to the gaugino coupling operators $g_2,\,g_Y$ 
respectively.
As mentioned earlier, a crucial role in soft leptogenesis is played by 
the $\widetilde N-\widetilde N^\dagger$ mixing to form the 
mass eigenstates
\bea
\widetilde{N}_{+}  &=&  
\frac{1}{\sqrt{2}}\left(e^{i\Phi/2}\widetilde{N}
+e^{-i\Phi/2}\widetilde{N}^{*}\right), \\ 
\widetilde{N}_{-}  &=& 
-\frac{i}{\sqrt{2}}(e^{i\Phi/2}\widetilde{N}
-e^{-i\Phi/2}\widetilde{N}^{*}),
\label{eq-08:mass_eigenstates}
\eea
where $\Phi \equiv\arg\left(B M\right)$ and the corresponding mass eigenvalues are 
$M_{\pm}^{2} = M^{2}+\widetilde{M}^{2}\pm\left|B M\right|$.
Without loss of generality, we can set $\Phi=0$,
which is equivalent to assigning the phases only to $A$ and $Y_{\alpha}$. 

It has been pointed out that the CP asymmetries for the decays of
$\widetilde{N}_{\pm}$ into scalars and fermions have opposite sign and
cancel each other at the leading
order~\cite{Grossman:2003jv,D'Ambrosio:2003wy,Fong:2009iu}, resulting
in a strongly suppressed total CP asymmetry $\sim {\cal
  O}(m_{soft}^3/M^3)$ where $m_{soft}$ is the scale of soft SUSY
breaking terms.  There are two possibilities that can rescue
leptogenesis: Firstly, thermal effects which break SUSY can spoil this
cancellation~\cite{Grossman:2003jv,D'Ambrosio:2003wy,Fong:2009iu}.
Secondly, non-superequilibration effects (see Section
\ref{sec-08:NSE}) which imply that lepton and slepton asymmetries
differ, can also spoil this cancellation~\cite{Fong:2010bv}.  The CP
asymmetries for the decays of $\widetilde{N}_{\pm}$ into scalars and
fermions are respectively given by
\bea
\label{eq-08:CP_s}
\epsilon_{\alpha}^s(T) & = & 
\bar \epsilon_\alpha \Delta^s(T),\\
\label{eq-08:CP_f}
\epsilon_{\alpha}^f(T) & = & 
-\bar \epsilon_\alpha \Delta^f(T),
\eea
where $\bar\epsilon_\alpha$ is the temperature independent term of
$\sim {\cal O}(m_{soft}/M)$ which contains contributions from the
self-energy correction, vertex correction and the interference between
the two.  In the limit $T \to 0$, we have $\Delta^s(T), \Delta^f(T)
\to 1/2$, and thus the inclusion of thermal effects and/or
non-superequilibration is mandatory to avoid the cancellation between
the asymmetries into scalars and fermions.

We can make a rough estimate of the scale relevant for soft
leptogenesis by requiring $|\bar\epsilon| \sim m_{soft}/M \gtrsim
10^{-6}$ which gives $M \lesssim 10^9$ GeV for $m_{soft} \sim 1$
TeV. Hence soft leptogenesis always happens in the temperature regime
where lepton flavour effects are relevant~\cite{Fong:2008mu}. In
general, the CP asymmetry from self-energy contribution requires $B
\ll m_{soft}$ to be resonantly enhanced. However it was shown in Ref. \cite{Fong:2010zu} 
that flavour effects can greatly enhance the efficiency and eventually $B
\sim m_{soft}$ is allowed.  The nice feature of
soft leptogenesis is that the tension with the gravitino problem gets
generically relaxed and, in the lower temperature window, is
completely avoided.

\subsection{Dirac leptogenesis}
\label{sec-08:dirac}
The extension of the SM with singlet neutrinos allows for 
two different ways for generating tiny neutrino masses.
The first one is the seesaw mechanism which has 
at least three attractive features:
\begin{itemize} \itemsep -1pt 
\item No extra symmetries (and, in particular, no global symmetries)
  have to be imposed.
\item The extreme lightness of neutrino masses is linked to the
  existence of a high scale of new physics, which is well
  motivated for various other reasons ({\it e.g.} gauge unification).
\item Lepton number is violated, which opens the way to leptogenesis.
\end{itemize}
The second way is to impose lepton number and give to the neutrinos
Dirac masses. A priori, one might think that all three attractive
features of the seesaw mechanism are lost. Indeed, one must usually
impose additional symmetries. But one can still construct natural
models where the tiny Yukawa couplings that are necessary for small
Dirac masses are related to a small breaking of a symmetry. What is
perhaps most surprising is the fact that leptogenesis could proceed
successfully even if neutrinos are Dirac particles, and lepton number
is not (perturbatively) broken~\cite{Akhmedov:1998qx,Dick:1999je}.
Such scenarios have been termed `Dirac leptogenesis'
\cite{Dick:1999je,Murayama:2002je,Boz:2004ga,Abel:2006hr,Cerdeno:2006ha,
  Thomas:2005rs,Thomas:2006gr,Chun:2008pg,Bechinger:2009qk,Chen:2011sb}.

An implementation of the idea is the following. A CP-violating decay
of a heavy particle can result in a non-zero lepton number for
LH particles, and an equal and opposite non-zero lepton
number for RH particles, so that the total lepton number is
zero. For the charged fermions of the SM, the Yukawa
interactions are fast enough that they quickly equilibrate the
LH and the RH particles, and the lepton number
stored in each chirality goes to zero. This is not true, however, for
Dirac neutrinos. The size of their Yukawa couplings is
$\lambda\lsim10^{-11}$, which means that equilibrium between the
lepton numbers stored in LH and RH neutrinos will
not be reached until the temperature falls well below the electroweak
breaking scale. To see this, note that the rate of the Yukawa
interactions, given by $\Gamma_\lambda\sim\lambda^2 T$, becomes
significant when it equals the expansion rate of the Universe, $H\sim
T^2/m_{\rm pl}$. Thus, the temperature of equilibration between
LH and RH neutrinos is $T\sim\lambda^2 m_{\rm
  pl}\sim (\lambda/10^{-11})^2\,$MeV, that is well below the
temperature when sphalerons, after having converted part of the
LH lepton asymmetry into a net baryon asymmetry, are switched
off.


A specific example of a supersymmetric model where Dirac neutrinos
arise naturally is presented in Ref. \cite{Murayama:2002je}. The
Majorana masses of the $N$-superfields are forbidden by $U(1)_L$. The
neutrino Yukawa couplings are forbidden by a $U(1)_N$ symmetry where,
among all the MSSM+N fields, only the $N$ superfields are charged. The
symmetry is spontaneously broken by the vacuum expectation value of a
scalar field $\chi$ that can naturally be at the weak scale,
$\langle\chi\rangle\sim v_u$. This breaking is communicated to the
MSSM+N via extra, vector-like lepton doublet fields, $\phi+\bar\phi$,
that have masses $M_\phi$ much larger than $v_u$.  Consequently, the
neutrino Yukawa couplings are suppressed by the small ratio
$\langle\chi\rangle/M_\phi$.  The CP violation arises in the decays of
the vector-like leptons, whereby $\Gamma(\phi\to
NH_u^c)\neq\Gamma(\bar\phi\to N^cH_u)$ and $\Gamma(\phi\to
L\chi)\neq\Gamma(\bar\phi\to L^c\chi^c)$. The resulting asymmetries in
$N$ and in $L$ are equal in magnitude and opposite in sign. Finally
note that the main phenomenological implication of Dirac leptogenesis
is the absence of any signal in neutrinoless double beta decays.

\subsection{Triplet scalar (type-II) leptogenesis}
\label{sec-08:type-II}

One can generate seesaw masses for the light neutrinos by tree-level
exchange of $SU(2)_L$-triplet scalars
$T$~\cite{Mohapatra:1980yp,Magg:1980ut,Schechter:1980gr,%
  Wetterich:1981bx,Lazarides:1980nt}. The relevant new terms in the
Lagrangian are
\be
\label{eq-08:ltrisca}
{\cal L}_T=-M_T^2|T|^2+\frac12\left([\lambda_L]_{\alpha\beta}\,\ell_\alpha
\ell_\beta T +M_T\lambda_\phi\,\phi\phi T^*+{\rm h.c.}\right).
\ee
Here, $M_T$ is a real mass parameter, $\lambda_L$ is a symmetric
$3\times3$ matrix of dimensionless, complex Yukawa couplings, and
$\lambda_\phi$ is a dimensionless complex coupling. 
Since this mechanism necessarily involves lepton number violation and
allows for new CP-violating phases, it is interesting to examine it in
the light of leptogenesis
\cite{Ma:1998dx,Chun:2000dr,Hambye:2000ui,Joshipura:2001ya,Hambye:2003ka,%
  D'Ambrosio:2004fz,Guo:2004mp,Antusch:2004xy,Antusch:2005tu,Chun:2005ms,%
  Hambye:2005tk,Chun:2006sp,Gu:2006wj,Sahu:2007uh,McDonald:2007ka,%
  Antusch:2007km,Chao:2007rm,Hallgren:2007nq,Frigerio:2008ai,Strumia:2008cf,Gu:2008yj,Calibbi:2009wk,
  Chen:2010uc,AristizabalSierra:2011ab,Arina:2011cu}.  One obvious
problem in this scenario is that, unlike singlet fermions, the triplet
scalars have gauge interactions that keep them close to thermal
equilibrium at temperatures $T\lsim10^{15}\>$ GeV. It turns out,
however, that successful leptogenesis is possible even at a much lower
temperature. This subsection is based in large part on
Ref. \cite{Hambye:2005tk} where further details and, in particular,
an explicit presentation of the relevant BE can be found.

The CP asymmetry that is induced by the triplet scalar decays is
defined as follows:
\be
\epsilon_T\equiv2\frac{\Gamma(\bar T\to \ell\ell)-\Gamma(T\to
  \bar\ell\bar\ell)}{\Gamma_T+\Gamma_{\bar T}},
\ee
where the overall factor of 2 comes because the triplet scalar decay
produces two (anti)leptons.

To calculate $\epsilon_T$, one should use the Lagrangian in
\eqn{eq-08:ltrisca}. While a single triplet is enough to produce three
light massive neutrinos, there is a problem in leptogenesis if indeed
this is the only source of neutrinos masses: The asymmetry is
generated only at higher loops and in unacceptably small.  It is still
possible to produce the required lepton asymmetry from a single
triplet scalar decays if there are additional sources for neutrino
masses, such as type I, type III, or type II contributions from
additional triplet scalars. Define $m_{\rm II}$ ($m_{\rm I}$) as the
part of the light neutrino mass matrix that comes (does not come) from
the contributions of the triplet scalar responsible for $\epsilon_T$:
\be
m=m_{\rm II}+m_{\rm I}.
\ee
Then, assuming that the particles  exchanged to produce
$m_{\rm I}$ are all heavier than $T$, we obtain the the CP asymmetry 
\be
\epsilon_T=\frac{1}{4\pi}\frac{M_T}{v_u^2}\sqrt{B_LB_H}\,\frac{{\rm
    Im}[{\rm Tr}(m_{\rm II}^\dagger m_{\rm I})]}{{\rm Tr}(m_{\rm
    II}^\dagger m_{\rm II})},
\ee
where $B_L$ ($B_H)$ is the tree-level branching ratio to leptons
(Higgs doublets). If these are the only decay modes, {\it i.e.}
$B_L+B_H=1$, then $B_L/B_H={\rm
  Tr}(\lambda_L\lambda_L^\dagger)/(\lambda_H\lambda_H^\dagger)$, and 
there is an upper bound on the asymmetry:
\be
\label{uppertri}
|\epsilon_T|\leq\frac{1}{4\pi}\frac{M_T}{v_u^2}\sqrt{B_LB_H\sum_i
  m_{\nu_i}^2}.
\ee
Note that, unlike the singlet fermion case, $|\epsilon_T|$ increases
with larger $m_{\nu_i}$.

As concerns the efficiency factor, it can be close to maximal,
$\eta\sim1$, in spite of the fact that the gauge interactions tend to
maintain the triplet abundance very close to thermal
equilibrium. There are two  necessary conditions that have to be
fulfilled by the  decay rates $T\to\bar\ell\bar\ell$ and
  $T\to\phi\phi$ in order that this will happen \cite{Hambye:2005tk}:
\begin{enumerate} \itemsep -1pt
\item One of the two decay rates is faster than the $T\bar T$
  annihilation rate.
\item The other decay mode is slower than the expansion rate of the
    Universe.
\end{enumerate}
The first condition guarantees that gauge scatterings are ineffective:
the triplets decay before annihilating. The second condition
guarantees that the fast decays do not washout strongly  the
lepton asymmetry: lepton number is violated only by the simultaneous
presence of $T\to\bar\ell\bar\ell$ and $T\to\phi\phi$.

Combining a calculation of $\eta$ with the upper bound on the CP
asymmetry (\ref{uppertri}), successful leptogenesis implies a lower
bound on the triplet mass $M_T$ varying between $10^9$ GeV and
$10^{12}$ GeV, depending on the relative weight of $m_{\rm II}$ and
$m_{\rm I}$ in the light neutrino mass.

Interestingly, in the supersymmetric framework, ``soft leptogenesis'' 
(see Section \ref{sec-08:soft}) 
can be successful even with the minimal set of extra fields -- a
single $T+\bar T$ -- that generates both neutrino masses and the lepton
asymmetry \cite{D'Ambrosio:2004fz,Chun:2005ms}.

\subsection{Triplet fermion (type-III) leptogenesis}
\label{sec-08:type-III}

One can generate neutrino masses by the tree level exchange of
$SU(2)_L$-triplet fermions $T_i^a$~\cite{Foot:1988aq,Ma:1998dn,Ma:2002pf}
($i$ denotes a heavy mass eigenstate while $a$ is an $SU(2)_L$
index) with the Lagrangian
\be
{\cal L}_{T^a}=[\lambda_T]_{\alpha
  k}\tau^a_{\rho\sigma}\ell_\alpha^\rho \phi^\sigma T_k^a-\frac12
  M_iT_i^aT_i^a+{\rm h.c.}.
\ee
Here $\tau^a$ are the Pauli matrices, $M_i$ are real mass parameters 
and $\lambda_T$ is a $3\times3$ matrix of complex Yukawa couplings.

This mechanism necessarily involves lepton number violation,
and allows for new CP-violating phases so we should examine it as a
possible source of leptogenesis~\cite{Brahmachari:2001bv,Hambye:2003rt,
Strumia:2008cf,Blanchet:2008cj,
Blanchet:2008ga,Patra:2010ks,AristizabalSierra:2010mv,Kannike:2011fx}.  
This subsection is based in large part on 
Ref.~\cite{Hambye:2003rt} 
where further details and  the relevant BE can be found.

As concerns neutrino masses, all the qualitative features are very
similar to the singlet fermion case. As concerns leptogenesis there
are, however, qualitative and quantitative differences.  With regard
to the CP asymmetry from the lightest triplet fermion decay, the
relative sign between the vertex and  self-energy
loop contributions is opposite to that of the singlet fermion
case. Consequently, in the limit of strong hierarchy in the heavy
fermion masses, the asymmetry in triplet decays is three times smaller than
in the decays of the singlets.  On the other hand, since the triplet has three
components, the ratio between the final baryon asymmetry and
$\epsilon\eta$ is three times bigger. The decay rate of the heavy fermion
is the same in both cases. This, however, means that the thermally
averaged decay rate is three times bigger for the triplet, as is the
on-shell part of the $\Delta L=2$ scattering rate.

A significant qualitative difference arises from the fact that the
triplet has gauge interactions. The effect on the washout factor
$\eta$ is particularly significant for $\tilde m\ll10^{-3}\>$eV, the
so-called ``weak washout regime'' (note that this name is
inappropriate for triplet fermions). The gauge interactions still
drive the triplet abundance close to thermal equilibrium. A relic
fraction of the triplet fermions survives. The decays of these relic
triplets produce a baryon asymmetry, with
\be
\eta\approx M_1/10^{13}\ {\rm GeV}\ \ \ ({\rm for}\ \tilde
m\ll10^{-3}\ {\rm eV}).
\ee
The strong dependence on $M_1$ results from the fact that the
expansion rate of the Universe is slower at lower temperatures.
On the other hand, for $\tilde m\gg10^{-3}\>$ eV, the Yukawa
interactions keep the heavy fermion abundance
close to thermal equilibrium, so the difference in $\eta$ between the
singlet and triplet case is only ${\cal O}(1)$.
Ignoring flavour effects, and assuming strong hierarchy between the
heavy fermions,  Ref. \cite{Hambye:2003ka} obtained 
the lower bound 
\be
M_1\gsim1.5\times10^{10}\ {\rm GeV}\,. 
\ee
When the triplet fermion scenario is incorporated in a supersymmetric
framework, and the soft breaking terms do not play a significant role,
the modifications to the above analysis is by factors of ${\cal O}(1)$.

\vspace{1cm}
\section{Conclusions} 
\label{sec-08:conclusions}

During the last few decades, a large set of experiments involving
solar, atmospheric, reactor and accelerator neutrinos have converged
to establish that the neutrinos are massive. The seesaw mechanism
extends the Standard Model in a way that allows neutrino masses, and
it provides a nice explanation of the suppression of the neutrino
masses with respect to the electroweak breaking scale.  Furthermore,
without any addition or modification, it can also account for the
observed baryon asymmetry of the Universe.  The possibility of giving
an explanation of two apparently unrelated experimental facts --
neutrino masses and the baryon asymmetry -- within a single framework
that is a natural extension of the Standard Model, together with the
remarkable `coincidence' that the same neutrino mass scale suggested
by neutrino oscillation data is also optimal for leptogenesis, make
the idea that baryogenesis occurs through leptogenesis a very
attractive one.

Leptogenesis can be quantitatively successful without any fine-tuning
of the seesaw parameters. Yet, in the non-supersymmetric seesaw
framework, a fine-tuning problem arises due to the large corrections
to the mass-squared parameter of the Higgs potential that are
proportional to the heavy Majorana neutrino masses. Supersymmetry can
cure this problem, avoiding the necessity of fine tuning; however, it
brings in the gravitino problem~\cite{Khlopov:1984pf} that requires a
low reheat temperature after inflation, in conflict with generic
leptogenesis models.  Thus, constructing a fully satisfactory
theoretical framework that implements leptogenesis within the seesaw
framework is not a straightforward task.

From the experimental side, the obvious question to ask is if it is
possible to test whether the baryon asymmetry has been really produced
through leptogenesis. Unfortunately it seems impossible that any
direct test can be performed.  To establish leptogenesis
experimentally, we need to produce the heavy Majorana neutrinos and
measure the CP asymmetry in their decays. However, in the most natural
seesaw scenarios, these states are simply too heavy to be produced,
while if they are light, then their Yukawa couplings must be very
tiny, again preventing any chance of direct measurements.

Lacking the possibility of a direct proof, experiments can still
provide circumstantial evidence in support of leptogenesis by
establishing that (some of) the Sakharov conditions for leptogenesis
are realized in nature.
Planned neutrinoless double beta decay ($0\nu\beta\beta$) experiments
(GERDA~\cite{GERDA}, MAJORANA~\cite{Aalseth:2004yt},
CUORE~\cite{Ardito:2005ar}) aim at a sensitivity to the effective
$0\nu\beta\beta$ neutrino mass in the few $\times$ 10 meV range.  If
they succeed in establishing the Majorana nature of the light
neutrinos, this will strengthen our confidence that the seesaw
mechanism is at the origin of the neutrino masses and, most
importantly, will establish that the first Sakharov condition for the
dynamical generation of a lepton asymmetry ($L$ violation)
is realized.
Proposed SuperBeam facilities~\cite{Autin:2000mn,Mezzetto:2003mm} and
second generation off-axis SuperBeam experiments
(T2HK~\cite{Itow:2001ee}, NO$\nu$A~\cite{Ayres:2004js}) can discover
CP violation in the leptonic sector. These experiments can only probe
the Dirac phase of the neutrino mixing matrix. They cannot probe
the Majorana low energy or the high energy phases, but the
important point is that they can establish that the second Sakharov
condition for the dynamical generation of a lepton asymmetry is
satisfied.
As regards the third condition, that is that the heavy neutrino decays
occurred out of thermal equilibrium, it might seem the most difficult
one to test experimentally. In reality the opposite is true, and in
fact we already know that an absolute neutrino mass scale of the order
of the solar or atmospheric mass differences is perfectly compatible
with sufficiently out of equilibrium heavy neutrinos decays.

Given that we do not know how to prove that leptogenesis is the
correct theory, we might ask if there is any chance to falsify it.
Indeed, future neutrino experiments could weaken the case for
leptogenesis, or even falsify it, mainly by establishing that the
seesaw mechanism is not responsible for the observed neutrino masses.
By itself, failure in revealing signals of $0\nu\beta\beta$ decays
will not disprove leptogenesis. Indeed, with normal neutrino mass
hierarchy one expects that the rates of lepton-number-violating
processes are below experimental sensitivity.  However, if neutrinos
masses are quasi-degenerate or inversely hierarchical, and future
measurements of the oscillation parameters will not fluctuate too much
away from the present best fit values, the most sensitive
$0\nu\beta\beta$ decay experiments scheduled for the near future
should be able to detect a signal \cite{Strumia:2006db}.  If instead
the limit on $|m_{\beta\beta}|$ is pushed below $\sim 10\,$meV (a
quite challenging task), this would suggest that either the mass
hierarchy is normal, or neutrinos are not Majorana particles.  The
latter possibility would disprove the seesaw model and standard
leptogenesis. Thus, determining the order of the neutrino mass spectrum
is extremely important to shed light on the connection between
$0\nu\beta\beta$ decay experiments and leptogenesis.
In summary, if it is established that the neutrino mass hierarchy is
inverted and at the same time no signal of $0\nu\beta\beta$ decays is
detected at a level $|m_{ee}|\lsim 10\,$meV, one could conclude that
the seesaw is not at the origin of the neutrino masses, and that
(standard) leptogenesis is not the correct explanation of the baryon
asymmetry.
As concerns CP violation, a failure in detecting leptonic CP violation
will not weaken the case for leptogenesis in a significant way.
Instead, it would mean that the Dirac CP phase is small enough to
render $L$-conserving CP-violating effects unobservable.

Finally, the CERN LHC has the capability of providing information that
is relevant to leptogenesis, since it can play a fundamental role in
establishing that the origin of the neutrino masses is not due to the
seesaw mechanism, thus leaving no strong motivation for leptogenesis.
This may happen in several different ways. For example (assuming that
the related new physics is discovered), the LHC will be able to test
if the detailed phenomenology of any of the following models is
compatible with an explanation of the observed pattern of neutrino
masses and mixing angles: supersymmetric R-parity violating couplings
and/or L-violating bilinear
terms~\cite{deCampos:2007bn,Allanach:2007vi};
leptoquarks~\cite{Mahanta:1999xd,AristizabalSierra:2007nf}; triplet
Higgses~\cite{Garayoa:2007fw,Kadastik:2007yd}; new scalar particles of
the type predicted in the Zee-Babu~\cite{Zee:1985id,Babu:1988ki} types
of models~\cite{AristizabalSierra:2006gb,Chen:2007dc,Nebot:2007bc}.
It is conceivable that such discoveries can eventually exclude the
seesaw mechanism and rule out leptogenesis.


\bibliography{fong}{}
\bibliographystyle{unsrt}

  \end{document}